\documentclass[12pt,a4paper]{article} 
\usepackage[utf8]{inputenc}
\usepackage[T1]{fontenc}
\usepackage[colorlinks,citecolor=blue,urlcolor=blue,linkcolor=blue]{hyperref}
\usepackage{graphicx}
\usepackage{dcolumn}
\usepackage{bm}
\usepackage{amsmath}
\usepackage{amsfonts}

\usepackage{titlesec}
\titlespacing*{\section}{0pt}{2ex}{2ex}
\titlespacing*{\subsection}{0pt}{2ex}{2ex} 
\titlespacing*{\subsubsection}{0pt}{2ex}{2ex}

\titleformat*{\section}{\large\bfseries}
\titleformat*{\subsection}{\large\bfseries}
\titleformat*{\subsubsection}{\large\bfseries}
\titleformat*{\paragraph}{\large\bfseries}
\titleformat*{\subparagraph}{\large\bfseries}

\usepackage{amssymb}
\usepackage{mathrsfs}
\usepackage{braket}
\usepackage{bbm}
\usepackage{xcolor}
\definecolor{olive}{rgb}{0.3, 0.4, .1}
\definecolor{fore}{RGB}{249,242,215}
\definecolor{back}{RGB}{51,51,51}
\definecolor{title}{RGB}{255,0,90}
\definecolor{blackViolet}{RGB}{138,43,226}
\definecolor{dgreen}{rgb}{0.,0.6,0.}
\definecolor{gold}{rgb}{1.,0.84,0.}
\definecolor{JungleGreen}{cmyk}{0.99,0,0.52,0}
\definecolor{blackGreen}{cmyk}{0.85,0,0.33,0}
\definecolor{RawSienna}{cmyk}{0,0.72,1,0.45}
\definecolor{Magenta}{cmyk}{0,1,0,0}
\definecolor{wood}{RGB}{139,115,85}
\definecolor{dorange}{RGB}{255,127,0}
\definecolor{dolive}{RGB}{85,107,47}
\definecolor{drg}{RGB}{255,165,0}

\DeclareMathAlphabet{\mathpzc}{OT1}{pzc}{m}{it}

\usepackage{multirow}

\usepackage{tikz}
\usepackage{colortbl}
\usepackage{multimedia}
\usepackage{xcolor}

\usepackage{graphicx}
\usepackage{ragged2e}
\usepackage{array}
\usepackage{rotating}
\usepackage{longtable}
\usepackage{arydshln}
\usepackage{multirow}

\usepackage{simplewick}
\usepackage{dirtytalk}

\usepackage{mathtools}

\newcommand\tvecpre{{\mathpalette\raiseT\top}}
\newcommand\raiseT[2]{\raisebox{-0.4ex}{$#1#2$}}
\newcommand\tvec{\mkern-2mu^\tvecpre\!}
\newcommand{\tmat}{\mkern-2mu ^\top\!}

\usepackage{stmaryrd}
\usepackage{geometry}
\usepackage{cite}

\makeatletter
\newcommand\xLongLeftRightArrow[2][]{%
  \ext@arrow 0099{\LongLeftRightArrowfill@}{#1}{#2}}
\def\LongLeftRightArrowfill@{%
  \arrowfill@\Leftarrow\Relbar\Rightarrow}
\makeatother
\usepackage{amssymb} 

\usepackage{amsthm}

\newtheorem{theorem}{Theorem}[section]
\newtheorem{lemma}{Lemma}[section]
\newtheorem{corollary}{Corollary}[section]
\newtheorem{proposition}{Proposition}[section]
\newtheorem{definition}{Definition}[section]

\let\emptyset\varnothing

\usepackage{dsfont}

\begin{document}

\title{\normalsize{\textbf{Boundary values for the charge transferred during an electronic transition}} \\ \normalsize{\textbf{Insights from matrix analysis}}}%

\date{}

\maketitle
\vspace*{-2cm}

\noindent \begin{center}
\textit{by}\footnote{Affiliation of the authors is given in a dedicated section right before the bibliography.} Enzo Monino, Jérémy Morere, \textit{and} Thibaud Etienne
\end{center}

$\;$


 

\begin{abstract}
\noindent In this contribution we start by proving and generalizing a conjecture that has been established few decades ago, relating the value of the integral of the detachment/attachment density in two pictures — one accounting for transition-induced basis relaxation and one which does not account for such a relaxation. To this end, we show that it is possible to follow two ways: one combines Haynsworth and Courant-Fischer theorems with a corollary to Lidskii-Wielandt theorem, the other combines two twin theorems extending Cauchy's interlacing theorem, together with the abovementioned corollary to Lidskii-Wielandt theorem. These derivations allow us to provide an upper bound for the electronic charge that is effectively displaced during the molecular electronic transition from one electronic quantum state to another. This quantity can be regarded as the neat charge that has been transferred during the transition. Our derivations ultimately show that this boundary value can be determined from a simple singular value decomposition and at most two matrix trace-computing operations.   \\ $\;$ 

\end{abstract}

\section{Introduction}
\noindent Providing quantitative insights into the light-induced molecular electronic-structure reorganization is a challenging task that has seen plenty of contributions providing complementary pieces of information \cite{barca_excitation_2018,breuil_diagnosis_2019-2,garcia_evaluating_2013,guido_metric_2013,huet_general_2020,kimber_toward_2020,le_bahers_nature_2014,ciofini_through-space_2012,luzanov_interpretation_1980,mai_quantitative_2018,mewes_density-based_2019,plasser_new_2014,etienne_toward_2014,etienne_new_2014,le_bahers_qualitative_2011}. Those approaches are often coupled with a qualitative analysis for visualization purposes \cite{bappler_exciton_2014,martin_natural_2003,dreuw_single-reference_2005,head-gordon_analysis_1995,li_visualisation_2016,le_bahers_qualitative_2011,li_particle-hole_2016,li_particlehole_2015,luzanov_charge_1978,etienne_towards_2021,kimber_toward_2020}. In this regard, two pairs of one-electron reduced density functions have attracted the attention of the molecular electronic excited states community. The first pair is the depletion/accumulation density pair \cite{le_bahers_qualitative_2011}, obtained by separating the positive-valued and negative-valued density contributions to the one-electron reduced difference density — note that ``depletion'' and ``accumulation'' names are not really used, but the density functions are. The second pair is the detachment/attachment density pair \cite{head-gordon_analysis_1995,etienne_towards_2021}, obtained by a similarity transformation of the one-electron reduced difference density matrix performed in order to obtain and separate the negative and positive ``transition occupation numbers'' and to collect the related information into two one-electron reduced density matrices. These two matrices are called ``detachment'' and ``attachment'' density matrices — rigorous foundations for this methodology are provided in ref. \cite{etienne_towards_2021}. The densities corresponding to these density matrices are visualized and analyzed as a ``departure/arrival'' density map, so that they can show some overlap, hence it is a picture different from the depletion/accumulation one. However, some relationships exist between the two pictures — for instance, the detachment/attachment density integral value is an upper bound to the depletion/accumulation density integral value.

When introducing the detachment/attachment picture \cite{head-gordon_analysis_1995}, Head-Gordon and co-workers gave an interesting comment relatively to the integral of the detachment/attachment densities: when using the configuration interaction singles (CIS) method \cite{hirata_configuration_1999} — for which this integral is equal to unity in the native picture —, a post-transition account for a transition-induced basis relaxation results in adding a relaxation matrix to the ``unrelaxed'' one-body reduced difference density matrix. The authors postulated that incorporating such a relaxation effect into the analysis will, when producing the eigenvalues of the one-electron reduced difference density matrix, make ``all negative eigenvalues becoming more negative and all positive eigenvalues becoming more positive. From the interleaving theorem of symmetric matrices, (...) the promotion number for the relaxed CIS density rigorously satisfies $p \geq 1$''. In the previous quote, $p$ is the detachment/attachment density integral — called promotion number; we will use another symbol for it in this contribution —, and the ``interleaving theorem'' is another name for Cauchy's interlacing theorem (\textit{vide infra}). This result has, to our knowledge, never been actually proved. Therefore, we would like to show in this contribution that not only this statement is true, but it can be extended to more sophisticated electronic excited-state calculation methods such as time-dependent Hartree-Fock theory \cite{mclachlan_time-dependent_1964}, time-dependent density-functional theory \cite{casida_time-dependent_1995}, or the Bethe-Salpeter method \cite{blase_bethesalpeter_2018}.

On the other hand, in a paper published almost two decades ago \cite{furche_adiabatic_2002}, Furche and co-workers have mentioned, about the time-dependent density-functional theory one-electron reduced difference density matrix, that ``in analogy to the ground state KS scheme, $P$ would yield the exact density difference if the exact (time-dependent) exchange-correlation functional were used. This follows from the fact that the density computed from $P$ is identical to the functional derivative of the excitation energy with respect to a local external potential.'' Here, ``KS'' stands for ``Kohn-Sham'', and $P$ is the relaxed one-electron reduced difference density matrix — another symbol for it will be used in this contribution.

According to some relationships that will be derived in this paper, we will see that it is possible to provide a theoretical value for the upper bound to the detachment/attachment density integral, hence the light-induced transferred density integral — the charge transferred during the molecular electronic transition. 

Proofs to the linear-algebraic propositions that are reported in this contribution without providing a proof directly in the text can be found in refs \cite{horn_matrix_2012,bhatia_matrix_1997,axler_linear_2015,roman_advanced_2008,bhatia2007perturbation,bhatia2009positive}.


\section{Hypotheses, notations, and background}
 \label{sec:hypotheses}
\noindent In this paper, ``$\wedge$'' will denote the logical conjunction, and ``$\lor$'' will denote the non-exclusive logical disjunction. In what follows, $n$, $m$, $N$, $M$ and $L$ are non-zero natural numbers. In this contribution, we will deal with $N$-electron systems in the field of $M$ fixed nuclei. The $N$ electrons are described using an $L$-dimensional orthonormal family of real-valued one-body wavefunctions: $$\mathcal{B} \coloneqq \left(\varphi_i\right)_{i\in \llbracket 1,L\rrbracket}.$$ 
In this paper, zero column vectors with $n$ components will be denoted $\textbf{0}_{n\times 1}$ while $n\times n$ zero matrices will be denoted $\textbf{0}_n$. We also provide here a definition for the Rayleigh quotient:
\begin{definition}[Rayleigh quotient]\label{def:rayleigh} The {\normalfont Rayleigh quotient} for complex matrices and vectors is defined as follows:
\begin{align*}
 & {\normalfont \forall n \in \mathbb{N}^*, \forall \textbf{A}\in \mathbb{C}^{n\times n}, \forall \textbf{w}\in \mathbb{C}^{n \times 1}\textbackslash \left\lbrace \textbf{0}_{n\times 1}\right\rbrace,} \\ 
 & {\normalfont R_{\textbf{A}}(\textbf{w}) \coloneqq \dfrac{\textbf{w}^\dag \textbf{A}\textbf{w}}{\textbf{w}^\dag\textbf{w}},}
\end{align*}
where the ``$\dag$'' symbol denotes Hermitian conjugation. 
\end{definition}
\noindent We also give the following definitions, that will be useful in further sections:
  \begin{definition} An $n$--tuple $(a_1^\downarrow, \dotsc , a_n^\downarrow)$ of real numbers is given in decreasing order if
  \begin{equation*}
a_1^\downarrow \geq \cdots \geq a_n^\downarrow.
\end{equation*}
\end{definition}
\begin{definition}
An $n$--tuple $(a_1^\uparrow, \dotsc , a_n^\uparrow)$ of real numbers is given in increasing order if
\begin{equation*}
a_1^\uparrow \leq \cdots \leq a_n^\uparrow.
\end{equation*}
\end{definition}
\begin{definition}[Total sum of positive and negative eigenvalues]
Let {\normalfont$\textbf{A}$} be an $n \times n$ Hermitian matrix with $n$ eigenvalues, i.e., $\normalfont \sigma(\textbf{A}) = \left\lbrace \lambda _i(\textbf{A}) \, : \, i \in \llbracket 1,n\rrbracket \right\rbrace.$ The {\normalfont total sum of its positive eigenvalues} will be denoted ${\normalfont S_+(\textbf{A})}$, i.e.,
$$\normalfont S_+(\textbf{A}) \coloneqq \sum _{i=1}^n \mathrm{max}\left(\lambda _i(\textbf{A}),0\right),$$
and the {\normalfont total sum of its negative eigenvalues} will be denoted ${\normalfont S_-(\textbf{A})}$, i.e.,
$$\normalfont S_-(\textbf{A}) \coloneqq \sum _{i=1}^n \mathrm{min}\left(\lambda _i(\textbf{A}),0\right).$$
\end{definition}
\begin{definition}[Partial sum of positive and negative eigenvalues] Let $n$ be a non-zero natural integer strictly superior to one. Let ${\normalfont \mathcal{P}(\llbracket 1, n \rrbracket )}$ be the power set of $ \llbracket 1, n \rrbracket$ and let $I$ be an element of $ \mathcal{P}(\llbracket 1, n \rrbracket )\setminus \{\emptyset,\llbracket 1,n\rrbracket\}$. Let {\normalfont \textbf{A}} be an $n\times n$ Hermitian matrix, and $\left(a_1,\ldots, a_n\right)$ the $n$-tuple of its eigenvalues listed in any order. The {\normalfont partial sum of its positive eigenvalues} on $I$ will be denoted ${\normalfont S_+^I(\textbf{A})}$, i.e.,
$$\normalfont S_+^I(\textbf{A}) \coloneqq \sum _{i\in I} \mathrm{max}\left(\lambda _i(\textbf{A}),0\right),$$
and the {\normalfont partial sum of its negative eigenvalues} on $I$ will be denoted ${\normalfont S_-^I(\textbf{A})}$, i.e.,
$$\normalfont S_-^I(\textbf{A}) \coloneqq \sum _{i\in I} \mathrm{min}\left(\lambda _i(\textbf{A}),0\right).$$
\end{definition}
\begin{lemma}\label{lemma:S+IAleqS+A}
Let $n$ be a non-zero natural integer strictly superior to one. Let {\normalfont \textbf{A}} be an $n\times n$ Hermitian matrix. Then,
\begin{equation*}
\normalfont\forall I \in \mathcal{P}(\llbracket 1, n \rrbracket )\setminus \{\emptyset,\llbracket 1,n\rrbracket\}, \, \forall \alpha \in \mathbb{R}, \, S^I_+(\textbf{A}) \geq \alpha \, \Longrightarrow S_+(\textbf{A}) \geq \alpha .
\end{equation*}
\end{lemma}
\begin{proof}
Let $I$ be any element of $\mathcal{P}(\llbracket 1, n \rrbracket )\setminus \{\emptyset,\llbracket 1,n\rrbracket\}$. Let $\left\lbrace \lambda _1(\textbf{A}),\ldots , \lambda_n(\textbf{A})\right\rbrace$ be the spectrum of \textbf{A}. We can simultaneously assert
\begin{equation*}
\forall j \in \llbracket 1, n \rrbracket \setminus I, \,  \lambda _j(\textbf{A}) > 0 \Longleftrightarrow S_+^I(\textbf{A})+\mathrm{max}(\lambda_j(\textbf{A}),0) > S_+^I(\textbf{A})  
\end{equation*}
and
\begin{equation*}
\forall j \in \llbracket 1, n \rrbracket \setminus I, \,    \lambda _j(\textbf{A}) \leq 0 \Longleftrightarrow S_+^I(\textbf{A})+\mathrm{max}(\lambda_j(\textbf{A}),0) = S_+^I(\textbf{A})  .
\end{equation*}
Then,
\begin{equation*}
S_+^I(\textbf{A}) + \!\!\sum _{j\in \llbracket 1, n \rrbracket \setminus I} \!\!\mathrm{max}(\lambda_j(\textbf{A}),0) \geq S_+^I(\textbf{A}),
\end{equation*}
which means that $S_+(\textbf{A}) \geq S_+^I(\textbf{A)}$. Lemma \ref{lemma:S+IAleqS+A} directly follows.
\end{proof}
\begin{lemma}\label{lemma:S-IAgeqS-A}
Let $n$ be a non-zero natural integer strictly superior to one. Let ${\normalfont \mathcal{P}(\llbracket 1, n \rrbracket )}$ be the power set of $ \llbracket 1, n \rrbracket$. Let {\normalfont \textbf{A}} be an $n\times n$ Hermitian matrix. Then,
\begin{equation*}
\normalfont\forall I \in \mathcal{P}(\llbracket 1, n \rrbracket )\setminus \{\emptyset,\llbracket 1,n\rrbracket\}, \, \forall \alpha \in \mathbb{R}, \, S^I_-(\textbf{A}) \leq \alpha \, \Longrightarrow S_-(\textbf{A}) \leq \alpha .
\end{equation*}
\end{lemma}
\noindent Proof of Lemma \ref{lemma:S-IAgeqS-A} follows the exact same procedure as proof of Lemma \ref{lemma:S+IAleqS+A}.
\subsection{(Relaxed) difference density matrices}
\noindent We start this section by recalling two important lemmas from linear algebra:
\begin{lemma}\label{lemma:prodPD}
\textit{Let} $\displaystyle
{\normalfont \textbf{A}}$ \textit{be an $m\times n$ complex matrix and {\normalfont $\textbf{A}^\dag$} its adjoint. Then,}
\begin{equation*}
({\normalfont\textbf{A}\textbf{A}^\dag} \succeq 0) \, \wedge \,
({\normalfont\textbf{A}^\dag\textbf{A}} \succeq 0),
\end{equation*}
{\it i.e., the product of a matrix by its adjoint is a positive semi-definite matrix.}
\end{lemma}
\begin{proof}
The non-zero eigenvalues of both matrices are squared singular values of $\textbf{A}$.
\end{proof}
\begin{lemma}\label{lemma:sumPD}
\textit{Let} $\displaystyle
{\normalfont \textbf{A}}$ \textit{and} {\normalfont $\textbf{B}$} \textit{be two $n\times n$ complex matrices. Then,}
\begin{equation*}
\normalfont (\textbf{A} \succeq 0 )\,\land \, (\textbf{B} \succeq 0) \;\Longrightarrow \; \left(\textbf{A}+\textbf{B} \succeq 0\right),
\end{equation*}
{\it i.e., the sum of two positive semi-definite matrices is a positive semi-definite matrix.}
\end{lemma}
\begin{proof}
This simply follows from the definition of positive semi-definiteness for a matrix.
\end{proof}
\noindent In the case of excited-state calculation methods mentioned in the introduction section, the relaxed one-electron reduced difference density matrix $\bm{\gamma}^\Delta_{rlx}$ elements read, in the $\mathcal{B}$ ordered basis,
\begin{align*} 
\forall (r,s) \in \llbracket 1,L\rrbracket ^2, \, \left(\bm{\gamma}^\Delta_{rlx}\right)_{r,s} = \left\langle \psi _0 \left| \left[\hat{T}^\dag , \left[ \hat{s}^\dag\hat{r},\hat{T}\right]\right] + \left[ \hat{s}^\dag\hat{r},\hat{Z}\right] \right|\psi _0 \right\rangle
\end{align*}
where $\psi _0$ is a one-determinant wavefunction, and
\begin{align*}
\hat{T} &= \sum _{i=1}^N \sum _{a=N+1}^L \textbf{x}_{ia}\hat{a}^\dag\hat{i} - \textbf{y}_{ia}\hat{i}^\dag\hat{a},\\
\hat{Z} &= \sum _{i=1}^N \sum _{a=N+1}^L \textbf{z}_{ia}\hat{a}^\dag \hat{i} - \textbf{z}_{ia} \hat{i}^\dag\hat{a}.
\end{align*}
In the expression of $\hat{T}$ and $\hat{Z}$, we have that $\hat{a}^\dag$ and $\hat{i}^\dag$ respectively denote the second-quantization creation operators corresponding to the $a^\mathrm{th}$  and $i^\mathrm{th}$ spinorbitals, while $\hat{a}$ and $\hat{i}$ denote the second-quantization annihilation operators corresponding to the $a^\mathrm{th}$ and $i^\mathrm{th}$ spinorbitals. The \textbf{x}, \textbf{y}, and \textbf{z} vectors all belong to $\mathbb{R}^{[N \times (L-N)]\times 1}$. We can recast their vector components into matrices as follows: For every $(i,a)$ in $\llbracket 1,N\rrbracket \times \llbracket N+ 1,L\rrbracket$, we set
\begin{align*}
\left(\textbf{X}\right)_{i,a-N} &\coloneqq \textbf{x}_{ia},\\
\left(\textbf{Y}\right)_{i,a-N} &\coloneqq \textbf{y}_{ia}, \\
\left(\textbf{Z}\right)_{i,a-N} &\coloneqq \textbf{z}_{ia},
\end{align*}
and, using rules recapitulated in refs. \cite{etienne_comprehensive_2021,etienne_towards_2021}, we find that the relaxed one-body reduced difference density matrix is the simple sum of two matrices
 \begin{equation} \label{eq:gammarlx=gammadelta+gammaz}
 \bm{\gamma}^\Delta + \bm{\gamma}^\mathrm{Z} \eqqcolon \bm{\gamma}^\Delta_{rlx} \in \mathbb{R}^{L\times L}
 \end{equation}
 where the unrelaxed one-body reduced difference density matrix, $\bm{\gamma}^\Delta$, is block-diagonal:
 \begin{equation} \label{eq:DeltaAB}
 \bm{\gamma}^\Delta :=  \left(\textbf{C}_1\oplus \textbf{C}_2\right)
 \end{equation}
 with $\displaystyle -\textbf{C}_1 \coloneqq \textbf{XX}\tmat + \textbf{YY}\tmat\;\mathrm{and}\;
 \textbf{C}_2 \coloneqq \textbf{X}\tmat\textbf{X} + \textbf{Y}\tmat\textbf{Y}.$ The case of CIS mentioned in the introduction of this article is simply obtained when the $\textbf{Y}$ matrix vanishes — a complete comment on the spectral decomposition of the unrelaxed CIS one-electron reduced difference density matrix in terms of the singular values and singular vectors of $\textbf{X}$ is given in Ref. \cite{etienne_auxiliary_2025}. The $\bm{\gamma}^{\mathrm{Z}}$ matrix in \eqref{eq:gammarlx=gammadelta+gammaz} is an $L\times L$ symmetric matrix, sum of two nilpotent matrices:
\begin{equation}\label{eq:gammaZ}
\bm{\gamma}^\mathrm{Z} = \begin{pmatrix} \textbf{0}_N & \textbf{Z} \\ \textbf{Z}\tmat & \textbf{0}_{L-N} \end{pmatrix}
\end{equation}
where $\textbf{0}_N$ and $\textbf{0}_{L-N}$ are the $N\times N$ and the $(L-N)\times (L-N$) zero matrices, respectively. In \eqref{eq:DeltaAB}, the $\textbf{C}_1$ and $\textbf{C}_2$ matrices are sums of normal matrices, hence they are both symmetric. According to lemmas \ref{lemma:prodPD} and \ref{lemma:sumPD}, $\textbf{C}_1$ is negative semidefinite while $\textbf{C}_2$ is positive semidefinite:
 \begin{align}
 0 &\succeq \textbf{C}_1 \in \mathbb{R}^{N\times N},\label{eq:Anegdef}\\
 0 &\preceq \textbf{C}_2 \in \mathbb{R}^{(L-N)\times (L-N)}\label{eq:Bposdef}
 \end{align}
 and $\textbf{C}_1$ and $\textbf{C}_2$ have the same trace, up to a sign:
 \begin{equation}\label{eq:vartheta}
 -\mathrm{tr}(\textbf{C}_1) = \mathrm{tr}(\textbf{C}_2).
 \end{equation}
 This number is of such importance in this contribution that it will be given a label using the $\vartheta$ symbol:
 \begin{equation*}
 \vartheta \coloneqq S_+(\bm{\gamma}^\Delta).
 \end{equation*}
 \subsection{Electronic transitions: density pictures}
 \noindent Let $\bm{\Delta}$ be a one-body reduced difference density matrix depicting the passage of an $N$-electron system from a given ``state'' to another, with a fixed geometry of the $M$ nuclei, and in the same basis. This passage can, for instance, be the electronic transition in the unrelaxed (respectively, relaxed) picture, and the $\bm{\Delta}$ matrix is equal to $\bm{\gamma}^\Delta$ (respectively, $\bm{\gamma}^\Delta_{rlx}$). Alternatively, one can also use a one-body reduced difference density matrix to represent the passage from the unrelaxed to the relaxed picture of the electronic transition, and $\bm{\Delta}$ is then simply equal to $\bm{\gamma}^\mathrm{Z}$. In any of these three cases, the $\bm{\Delta}$ matrix is symmetric.

Once this ``departure/arrival'' one-body reduced difference density matrix has been identified, one can use the reduced density matrix theory to find the basis in which the matrix representation of this ``passage'' is diagonal. Such an orthogonal diagonalization produces the eigenvalues of the chosen $\bm{\Delta}$ matrix — eigenvalues ordering is arbitrary here:
\begin{equation*}
\exists (\textbf{P},\textbf{Q})\in \mathbb{R}^{L\times L} \times \mathbb{R}^{L\times L},\, \textbf{P}\tmat \bm{\Delta} \textbf{P} = \textbf{Q}
\end{equation*} 
with $\textbf{Q} = \mathrm{diag}\left(\lambda_i(\bm{\Delta})\right)_{i\in \llbracket 1,L\rrbracket}.$ The diagonal entries of $\textbf{Q}$ are called difference occupation numbers and are then used to create two arrays of positive numbers, $\textbf{Q}_+$ and $\textbf{Q}_-$, defined as follows:
\begin{align*}
2\textbf{Q}_+ &:= \mathrm{diag}\left( |\lambda_i(\bm{\Delta})|+\lambda_i(\bm{\Delta}) \right)_{i\in \llbracket 1,L\rrbracket}, \\
2\textbf{Q}_- &:= \mathrm{diag}\left( |\lambda_i(\bm{\Delta})|-\lambda_i(\bm{\Delta}) \right)_{i\in \llbracket 1,L\rrbracket}.
\end{align*}
The original orthogonal transformation matrix $\textbf{P}$ is then used to backtransform $\textbf{Q}_-$ and $\textbf{Q}_+$ in $\mathcal{B}$, respectively producing the so-called one-particle detachment ($\textbf{Q}_\mathrm{D}$) and attachment ($\textbf{Q}_\mathrm{A}$) density matrices in $\mathcal{B}$:
\begin{align*}
\textbf{Q}_\mathrm{D} \coloneqq \textbf{P}\textbf{Q}_-\textbf{P}\tmat \; \mathrm{and} \;
\textbf{Q}_\mathrm{A} \coloneqq \textbf{P}\textbf{Q}_+\textbf{P}\tmat, 
\end{align*}
with the property that
\begin{equation}\label{eq:A-D=Delta}
\textbf{Q}_\mathrm{A}-\textbf{Q}_\mathrm{D}= \bm{\Delta}. 
\end{equation}
Such a procedure applied to the unrelaxed difference density matrix produces the unrelaxed detachment and attachment density matrices that will be written $ \bm{\gamma}^d$ and $ \bm{\gamma}^a$, respectively. If the relaxed one-body reduced difference density matrix is diagonalized instead, the relaxed detachment and attachment density matrices will be written $ \bm{\gamma}^d_{rlx}$ and $ \bm{\gamma}^a_{rlx}$, respectively. Finally, if we are interested in deriving a detachment/attachment picture of the unrelaxed-relaxed transition process, it is the $\bm{\gamma}^\mathrm{Z}$ matrix that should be diagonalized to produce the relaxation detachment ($\bm{\gamma}^d_z$) and attachment ($\bm{\gamma}^a_z$) density matrices. According to \eqref{eq:Anegdef} and \eqref{eq:Bposdef}, we have that
 \begin{align*}
 \bm{\gamma}^d = \left(-\textbf{C}_1\right)\oplus \textbf{0}_{L-N} \; \mathrm{and} \;
 \bm{\gamma}^a = \textbf{0}_N \oplus \textbf{C}_2.
 \end{align*}
 We also notice that, according to \eqref{eq:vartheta}, $\bm{\gamma}^\Delta$, $\bm{\gamma}^\Delta_{rlx}$, and $\bm{\gamma}^\mathrm{Z}$ are all zero-trace. Therefore,
 $$S_+(\bm{\Delta}) = - S_-(\bm{\Delta}).$$
 Using \eqref{eq:A-D=Delta}, we deduce that, for a given $\bm{\Delta}$ (i.e., either $\bm{\gamma}^\Delta$, $\bm{\gamma}^\Delta_{rlx}$, or $\bm{\gamma}^\mathrm{Z}$), the trace of the corresponding detachment density matrix will be equal to the trace of the corresponding attachment density matrix:
 \begin{equation}\label{eq:trgammadgammaa}
 \mathrm{tr}(\textbf{Q}_\mathrm{D}) = \mathrm{tr}(\textbf{Q}_\mathrm{A}),
 \end{equation}
 and this number is equal to $S_+(\bm{\Delta})$. For $\bm{\gamma}^\Delta$, this number has already been labeled $\vartheta$; for $\bm{\gamma}^\Delta_{rlx}$, this number will be labeled $\vartheta^{rlx}$; for $\bm{\gamma}^\mathrm{Z}$, this number will be labeled $\vartheta^\mathrm{Z}$. 

Any of these detachment/attachment density matrices, or the one-body reduced difference density matrix, can be used to build the corresponding detachment/attachment one-particle density function or one-electron difference density function, respectively. For instance, in the unrelaxed picture, these three functions read, in the $\mathcal{B}$ ordered basis,
\begin{align*}
n_\omega \, : \, S_4 \, &\longrightarrow \, \mathcal{I} \\
\textbf{s} \, & \longmapsto \, n_\omega(\textbf{s}) = \sum _{r=1}^L\sum_{s=1}^L \left(\bm{\gamma}^\omega\right)_{r,s} \varphi _r(\textbf{s})\varphi _s(\textbf{s})
\end{align*}
where $S_4 \coloneqq \mathbb{R}^3 \times \llbracket 1, 2\rrbracket,$ and $\omega$ can be either $d$ or $a$ (in which case $\mathcal{I} = \mathbb{R}_+$) or $\Delta$ (in which case $\mathcal{I} = \mathbb{R}$). The integral over $S_4$ of $n_d$ and $n_a$ is equal to the trace of the unrelaxed detachment and attachment density matrices:
\begin{equation*}
\int _{S_4} \mathrm{d}\textbf{s} \, n_d(\textbf{s}) = \int _{S_4} \mathrm{d}\textbf{s} \, n_a(\textbf{s}) = \vartheta,
\end{equation*}
and similarly for $\vartheta^{rlx}$ and $\vartheta^{\mathrm{Z}}$. Note that this is not exactly a definite integral, but rather a sum over the discrete values for the spin projection, and an integral over $\mathbb{R}^3$.

\noindent We introduce another pair of density functions:
\begin{align*}
n_+ \, : \, S_4 \, &\longrightarrow \, \mathbb{R}_+ \\
\textbf{s} \, & \longmapsto \, n_+(\textbf{s}) = \dfrac{\left|n_\Delta(\textbf{s})\right| + n_\Delta(\textbf{s})}{2}, \\
n_- \, : \, S_4 \, &\longrightarrow \, \mathbb{R}_+ \\
\textbf{s} \, & \longmapsto \, n_-(\textbf{s}) = \dfrac{\left|n_\Delta(\textbf{s})\right| - n_\Delta(\textbf{s})}{2},
\end{align*}
The first one keeps only the positive entries of $n_\Delta$ on $S_4$ — i.e., the positive contributions to the difference density —, while the second one keeps only the absolute value of the negative contributions to the difference density. We have
\begin{equation}\label{eq:intn+=intn-}
\int _{S_4} \mathrm{d}\textbf{s} \, n_+(\textbf{s}) = \int _{S_4} \mathrm{d}\textbf{s} \, n_-(\textbf{s}).
\end{equation}
Two numbers have then been introduced to assess the locality of the electronic transition in the unrelaxed and relaxed pictures. In the unrelaxed picture first, the dimensionless, normalized spatial overlap between the unrelaxed detachment and attachment densities is computed as \cite{etienne_toward_2014}
\begin{equation*}
\phi _S := \vartheta^{-1} \int _{S_4}\mathrm{d}\textbf{s}\, \sqrt{n_d(\textbf{s})n_a(\textbf{s})},
\end{equation*}
while the neat displaced charge, i.e., the charge transferred during the electronic transition when one takes the before/after ``bilan'' picture, reads \cite{le_bahers_qualitative_2011}
\begin{equation*}
0 \leq q^\mathrm{CT} \coloneqq \int _{S_4} \mathrm{d}\textbf{s} \, n_+(\textbf{s}) .
\end{equation*}
Due to \eqref{eq:intn+=intn-}, we have that
\begin{equation*}
q^\mathrm{CT} = \dfrac{1}{2}\int _{S_4} \mathrm{d}\textbf{s} \, |{n_\Delta(\textbf{s})}| .
\end{equation*}
From \eqref{eq:A-D=Delta}, one can show that $(n_a - n_d = n_\Delta)$. Hence, we deduce
\begin{equation*}
\forall \textbf{s}\in S_4,\, |{n_\Delta(\textbf{s})}| = |n_a(\textbf{s})-n_d(\textbf{s})| \leq |n_a(\textbf{s})+n_d(\textbf{s})|
\end{equation*}
since the values of $n_d$ and $n_a$ lie in $\mathbb{R}_+$. The same reason can be invoked for justifying the fact that $|n_a(\textbf{s})+n_d(\textbf{s})| = n_a(\textbf{s})+n_d(\textbf{s})$. This actually allows us to write another inequality, namely
\begin{equation*}
\dfrac{1}{2}\int _{S_4} \mathrm{d}\textbf{s} \, |{n_\Delta(\textbf{s})}| \leq \dfrac{1}{2}\left( \int _{S_4} \mathrm{d}\textbf{s} \, {n_a(\textbf{s})} + \int _{S_4} \mathrm{d}\textbf{s} \, {n_d(\textbf{s})}\right),
\end{equation*}
leading finally to
\begin{equation}\label{eq:chiphiU}
0 \leq q^\mathrm{CT} \leq \vartheta.
\end{equation}
Extension of $q^\mathrm{CT}$ to the relaxed picture is straightforward, and gives what we will write $q^\mathrm{CT}_{rlx}$ in what follows, with
\begin{equation}
\label{eq:chiphiR}
0 \leq q^\mathrm{CT}_{rlx} \leq \vartheta^{rlx}.
\end{equation}
This is due to the fact that \eqref{eq:chiphiU} was established from the structure of the objects used for the construction of the $q^\mathrm{CT}$ and $\vartheta$ numbers, independently from the numerical content of these objects.

 \section{Lower bound to the relaxed detachment/attachment trace}
 \label{sec:bound}


\subsection{Derivation using Haynsworth and Courant-Fischer theorems}
In this paper, we will have to use the theorem of subspace intersection — see Lemma 4.2.3 of Ref. \cite{horn_matrix_2012}. It asserts the following:
\begin{lemma}[Subspace intersection theorem]\label{lemma:subspaceintersection}
Let $V$ be a vector space. Let $V_1$ and $V_2$ be two subspaces of $V$. Then,
\begin{equation*}
\left(\mathrm{dim} V_1 + \mathrm{dim} V_2 > \mathrm{dim} V \right)\Longleftrightarrow \left\lbrace \normalfont \textbf{v}  \, : \, \left({\textbf{v}} \in V_1 \cap V_2\right) \wedge \textbf{v} \neq \mathrm{\textbf{0}}_V\right\rbrace \neq \emptyset.
\end{equation*}
\end{lemma}
\begin{definition}[Inertia of a complex matrix] Let n be a non-zero natural integer. \textit{Let} {\normalfont{\textbf{M}}} \textit{be a complex, square matrix of order $n$. The {\normalfont{inertia}} of} {\normalfont{\textbf{M}}}, noted {\normalfont{$\mathrm{In}(\textbf{M})$}}, is the ordered triple of natural integers
\begin{equation*}
\mathrm{In}({\normalfont{\textbf{M}}}) = (\pi_M,\nu_M,\delta_M),
\end{equation*}
\textit{where} $\pi_M$ \textit{is the number of eigenvalues of} {\normalfont{\textbf{M}}} \textit{with a strictly positive real part}, $\nu_M$ \textit{is the number of eigenvalues of} {\normalfont{\textbf{M}}} \textit{with a strictly negative real part, and} $\delta_M$ \textit{is the number of eigenvalues of} {\normalfont{\textbf{M}}} with a zero real part. In particular, if {\normalfont \textbf{M}} is Hermitian, $\pi_M$ {\normalfont(}respectively, $\nu_M)$ \textit{is the number of its strictly positive {\normalfont(}respectively, strictly negative{\normalfont)} eigenvalues.}
\end{definition}
\noindent Notice that, for a given matrix \textbf{M}, we can alternatively use either $(\pi_M, \nu_M,\delta_M)$ or $(\pi,\nu,\delta)$, or even $(\pi(\textbf{M}),\nu(\textbf{M}),\delta(\textbf{M}))$ in the text.

Before going further, we will also recall the definition of the Moore-Penrose pseudoinverse \cite{penrose_generalized_1955}, together with the definition of the generalized Schur complement \cite{carlson_generalization_1974}. 
\begin{definition}[Moore-Penrose pseudoinverse] Let n and m be two non-zero natural integers. Let {\normalfont{\textbf{M}}} be an $n\times m$ complex matrix. Its {\normalfont Moore-Penrose pseudoinverse}, written {\normalfont{$\textbf{M}^+$}}, is defined such that the following conditions are fulfilled:
\begin{align*}
\normalfont (\textbf{M}\textbf{M}^+)\textbf{M} = \normalfont\textbf{M} \; &\wedge \normalfont(\textbf{MM}^+)^\dag = \normalfont\textbf{MM}^+ ,\\
\normalfont (\textbf{M}^+\textbf{M})\textbf{M}^+ = \normalfont\textbf{M}^+ \; &\wedge \;
\normalfont (\textbf{M}^+\textbf{M})^\dag = \normalfont\textbf{M}^+\textbf{M}.
\end{align*}
Moreover, if {\normalfont \textbf{M}} is non-singular, then
\begin{equation*}
\normalfont \textbf{M}^+ = \textbf{M}^{-1}.
\end{equation*}
\end{definition}
\begin{definition}[Generalized Schur complement]  Let $n$ and $m$ be two non-zero natural integers such that $m<n$. Let {\normalfont \textbf{A}} be a matrix partitioned as
\begin{eqnarray} \label{eq:matrix_shape} \normalfont 
\textbf{A} = 
 \left(
  \begin{array}{ll}
   \textbf{B}      &  \textbf{D} \\
   \textbf{D}^\dag &  \textbf{C}  \\
  \end{array} 
 \right)
\end{eqnarray}
{with} ${\normalfont \textbf{B}\in\mathbb{C}^{m\times m}}$, ${\normalfont \textbf{C}\in\mathbb{C}^{(n-m)\times(n-m)}}$, and ${\normalfont\textbf{D}\in\mathbb{C}^{m\times(n-m)}}$. The {\normalfont generalized Schur complement of} {\normalfont \textbf{B}} {\normalfont in} {\normalfont\textbf{A}} is defined as
\begin{equation*}\normalfont
\textbf{A}/\textbf{B} \coloneqq \textbf{C} - \textbf{D}^\dag\textbf{B}^+\textbf{D}.
\end{equation*}
Similarly, the {\normalfont generalized Schur complement of} {\normalfont\textbf{C}} {\normalfont in} {\normalfont\textbf{A}} is defined as
\begin{equation*}\normalfont
\textbf{A}/\textbf{C} \coloneqq \textbf{B} - \textbf{D}\textbf{C}^+\textbf{D}^\dag.
\end{equation*}
\end{definition}
\noindent At some point in the paper, our task will be to compare different matrices based on their inertia. For this purpose, we first bring a definition of an order relationship between tuples of natural integers:
\begin{definition}[$\geq$, the order relationship between tuples of natural integers]\label{def:orderrelationship}  Let $k$ be a non-zero natural integer. The {\normalfont ``$\geq$''} {\normalfont order relationship between $k$--tuples of natural integers} is defined such that if $a \coloneqq (a_i)_{i\in\llbracket 1,k\rrbracket}$ and $b\coloneqq (b_i)_{i\in\llbracket 1,k\rrbracket}$ are two $k$--tuples of natural integers, 
\begin{equation*}
a \geq b \Longleftrightarrow \!\!\bigwedge_{i\in\llbracket 1, k \rrbracket}\! (a_i \geq b_i).
\end{equation*}
\end{definition}
\noindent In 1968, Emilie Haynsworth proved a theorem related to the inertia of partitioned Hermitian matrices \cite{haynsworth_determination_1968}. In 1974, with two co-authors she generalized this result and proved a theorem about the inertia additivity of Hermitian matrices:
\begin{theorem}[Generalized Haynsworth inertia additivity I] \label{thm:generalizedhaynsworth1} Let {\normalfont \textbf{A}} be a Hermitian matrix partitioned as in \eqref{eq:matrix_shape}. Then,
\begin{equation*} \normalfont
\mathrm{In}(\textbf{A}) \geq \mathrm{In}(\textbf{B}) + \mathrm{In}(\textbf{A}/\textbf{B}),
\end{equation*}
where the addition between two 3-tuples is defined elementwise.
\end{theorem} 
\noindent Haynsworth and co-authors proved Theorem \ref{thm:generalizedhaynsworth1} in Ref. \cite{carlson_generalization_1974}. A twin theorem can be given for the other submatrix:
\begin{theorem}[Generalized Haynsworth inertia additivity II] \label{thm:generalizedhaynsworth2} Let {\normalfont \textbf{A}} be a Hermitian matrix partitioned as in \eqref{eq:matrix_shape}. Then,
\begin{equation*} \normalfont
  \mathrm{In}(\textbf{A}) \geq \mathrm{In}(\textbf{C}) + \mathrm{In}(\textbf{A}/\textbf{C}).
\end{equation*} 
\end{theorem} 
\begin{proof}
We introduce the $m$- and $(n-m)$-dimensional identity matrices $\textbf{I}_m$ and $\textbf{I}_{(n-m)}$, and 
\begin{eqnarray*}
\textbf{J} \coloneqq
 \left(
  \begin{array}{ll}
   \textbf{0}_{m\times(n-m)}       &  \textbf{I}_{m} \\
   \textbf{I}_{(n-m)} &  \textbf{0}_{(n-m)\times m}  \\
  \end{array} 
 \right),
\end{eqnarray*}
where $\textbf{0}_{m\times(n-m)}$ and $\textbf{0}_{(n-m)\times m}$ are the $m\times (n-m)$ and $(n-m)\times m$ zero matrices, respectively. The \textbf{J} matrix is orthogonal, i.e., $\textbf{J}^{-1} = \textbf{J}\tmat$. We then use \textbf{J} as a change-of-basis matrix:
\begin{eqnarray*} 
\textbf{J}^{-1}\textbf{A}\textbf{J} = 
 \left(
  \begin{array}{ll}
   \textbf{C} &  \textbf{D}^\dag \\
   \textbf{D} &  \textbf{B}  \\
  \end{array} 
 \right)
 \eqqcolon \textbf{A}'.
\end{eqnarray*}
Applying Theorem \ref{thm:generalizedhaynsworth1} to $\textbf{A}'$, i.e.,
\begin{equation*}
\mathrm{In}(\textbf{A}') \geq \mathrm{In}(\textbf{C}) + \mathrm{In}({\textbf{A}'/\textbf{C}}),
\end{equation*}
and noticing that \textbf{A} and $\textbf{A}'$ matrices are similar and thus share the same spectrum, i.e., ($\sigma(\textbf{A}) = \sigma(\textbf{A}')$),
we deduce that they share the same inertia: $\mathrm{In}(\textbf{A}) = \mathrm{In}(\textbf{A}')$. Noticing that the generalized Schur complements of both \textbf{A} et $\textbf{A}'$ matrices are the same, i.e.
\begin{equation*}
\textbf{A}'/\textbf{B} = \textbf{C} - \textbf{D}^\dag\textbf{B}^+\textbf{D} = \textbf{A}/\textbf{B}
\quad \mathrm{and}\quad
\textbf{A}'/\textbf{C} = \textbf{B} - \textbf{D}\textbf{C}^+\textbf{D}^\dag = \textbf{A}/\textbf{C}
\end{equation*}
then completes the proof.
\end{proof}
\noindent In our paper we will need a corollary of theorems \ref{thm:generalizedhaynsworth1} and \ref{thm:generalizedhaynsworth2} that states
\begin{corollary}\label{cor:haynsworth} Let {\normalfont \textbf{A}} be a Hermitian matrix partitioned as in \eqref{eq:matrix_shape}. Then,
\begin{equation*}\normalfont
(\mathrm{In}(\textbf{A}) \geq \mathrm{In}(\textbf{B}) ) \wedge (\mathrm{In}(\textbf{A}) \geq \mathrm{In}(\textbf{C})).
\end{equation*}
\end{corollary}
\begin{proof}
Since the elements of the inertia of a matrix are natural integers, we have, using Definition \ref{def:orderrelationship},
\begin{equation*} 
\mathrm{In}(\textbf{B}) + \mathrm{In}(\textbf{A}/\textbf{B})) \geq \mathrm{In}(\textbf{B}). 
\end{equation*}
Hence,
\begin{equation*}
\mathrm{In}(\textbf{A}) \geq \mathrm{In}(\textbf{B}) + \mathrm{In}(\textbf{A}/\textbf{B}) \Longrightarrow \mathrm{In}(\textbf{A}) \geq \mathrm{In}(\textbf{B}).
\end{equation*}
A similar reasoning holds for the inequality of Theorem \ref{thm:generalizedhaynsworth2}.
\end{proof}
\noindent This leads us to formulate the following lemma:
\begin{lemma}\label{lemma:pioverpic}
\noindent Let $n$ be a non-zero natural integer strictly superior to one. Let {\normalfont{\textbf{A}}} \textit{be an} $n\times n$ \textit{Hermitian matrix partitioned as}
\begin{eqnarray*}\normalfont
\textbf{A} = 
 \left(
  \begin{array}{ll}
   \textbf{B}      &  \textbf{D} \\
   \textbf{D}^\dag &  \textbf{C}  \\
  \end{array} 
 \right)
\end{eqnarray*}
\textit{with}
\begin{equation*}\normalfont
0\succeq \textbf{B} \in \mathbb{C}^{m\times m} \quad \mathrm{and} \quad 0\preceq \textbf{C} \in \mathbb{C}^{(n-m)\times (n-m)}
\end{equation*}
\textit{and assuming} $1 \leq m < n$. Let $ (\pi,\nu,\delta)$, $ (0,\nu_B,\delta_B)$, and $  (\pi_C,0,\delta_C)$
be the three 3-tuples corresponding to the inertia of {\normalfont \textbf{A}}, {\normalfont\textbf{B}}, and {\normalfont \textbf{C}}, respectively. Then,
\begin{equation*}
(\nu_B\leq \nu)\wedge (\pi_C \leq \pi).
\end{equation*}
\end{lemma}
\begin{proof}
This is a direct consequence of Corollary \ref{cor:haynsworth}.
\end{proof}
\noindent The following lemma is more general:
\begin{lemma}\label{lemma:piinftonminusm}
Let $n$ be a non-zero natural integer strictly superior to one. Let {\normalfont{\textbf{A}}} \textit{be an} $n\times n$ \textit{Hermitian matrix}. Let {\normalfont($\pi, \nu$, $\delta$)} be the inertia of {\normalfont{\textbf{A}}}. Let $m$ be a natural integer strictly superior to zero and strictly inferior to $n$. Then,
\begin{equation*}
(\nu \leq m) \lor (\pi \leq n-m).
\end{equation*}
\end{lemma}
\begin{proof}(\textit{Reductio ad absurdum}). The total number of eigenvalues of a square matrix is equal to its dimension:
\begin{equation}\label{eq:equality_dimension}
\pi + \nu + \delta = n.
\end{equation}
Suppose that $(\pi > n-m) \land (\nu > m)$ is true. The consequence of this would be that $(\pi + \nu  > n)$, which is in contradiction with \eqref{eq:equality_dimension}. Moreover, we know from its syntax that the following proposition is always true:
\begin{equation*}
\neg((\nu>m) \land (\pi > n-m)) \Longleftrightarrow (\nu \leq m) \lor (\pi \leq n-m).
\end{equation*}
We know from the present proof that its left-hand side is true. We therefore conclude that its right-hand side is also true.
\end{proof}
\noindent In section 4.2 of Ref. \cite{horn_matrix_2012}, on the way of proving the so-called Courant-Fischer theorem — see Theorem \ref{theorem:courant-fischer} below —, we find the following two lemmas:
\begin{lemma}\label{lemma:subsetssupinf}
Let $f$ be a bounded real-valued function whose domain is $\mathcal{D}$. Let $\mathcal{D}_1$ and $\mathcal{D}_2$ be non-empty sets such that $\mathcal{D}_1\subseteq \mathcal{D}_2\subseteq \mathcal{D}$. Then, the following three inequalities are verified:
\begin{equation}\label{eq:subsetsupinf1}
\sup_{x \in \mathcal{D}_2}f(x) \geq \sup_{x \in \mathcal{D}_1}f(x) ,
\end{equation}
\begin{equation}\label{eq:subsetsupinf2}
\sup_{x \in \mathcal{D}_1}f(x) \geq \inf_{x \in \mathcal{D}_1}f(x) ,
\end{equation}
\begin{equation}\label{eq:subsetsupinf3}
\inf_{x \in \mathcal{D}_1}f(x) \geq \inf_{x \in \mathcal{D}_2}f(x).
\end{equation}
\end{lemma}
\begin{lemma}\label{lemma:infsup}
 Let $n$ be a non-zero natural integer. Let {\normalfont\textbf{A}} be an $n\times n$ Hermitian matrix. Let $j$ be a non-zero natural integer inferior or equal to $n$. Let $\mathcal{S}_j$ be the set of every $j$--dimensional subspaces of $\mathbb{C}^n$. We define the corresponding $\mathcal{S}_j^*$ set as 
 \begin{equation*}\normalfont
 \mathcal{S}_j^* \coloneqq \bigcup _{\mathcal{S} \in \mathcal{S}_j} \left(\mathcal{S}\setminus\left\lbrace \textbf{0}_{\mathcal{S}}\right\rbrace\right).
 \end{equation*}
Let $k$ be a non-zero natural integer inferior or equal to $n$. Then,
 \begin{equation}\normalfont\label{eq:pre-CF_1}
 \inf_{S\in \mathcal{S}_k^*}\sup_{{\normalfont \textbf{w}}\in S} R_{\textbf{A}}({\textbf{w}}) =  \min_{S\in \mathcal{S}_k^*}\max_{{\normalfont \textbf{w}}\in S} R_{\textbf{A}}(\normalfont{\textbf{w}}),
 \end{equation}
 and
  \begin{equation}\normalfont\label{eq:pre-CF_2}
 \sup_{S \in \mathcal{S}_{n-k+1}^*}\inf_{{\normalfont \textbf{w}}\in S} R_{\textbf{A}}(\normalfont{\textbf{w}}) =  \max_{S\in \mathcal{S}_{n-k+1}^*}\min_{{\normalfont \textbf{w}}\in S} R_{\textbf{A}}(\normalfont{\textbf{w}}).
 \end{equation}
\end{lemma}
\noindent In Lemma \ref{lemma:infsup}, $R$ denotes the Rayleigh quotient defined in Definition \ref{def:rayleigh}.
\begin{theorem}[Courant-Fischer]\label{theorem:courant-fischer} Let $n$ be a non-zero natural integer. Let {\normalfont\textbf{A}} be an $n\times n$ Hermitian matrix. Let 
\begin{equation*}\normalfont
 (\lambda ^\uparrow _1(\textbf{A}),\ldots, \lambda ^\uparrow _n(\textbf{A})) 
\end{equation*} 
be the $n$-tuple of its eigenvalues sorted in increasing order. Let $k$ be a non-zero natural integer inferior or equal to $n$. Let $\mathcal{S}_k^*$ and $\mathcal{S}_{n-k+1}^*$ be the two sets defined in {\normalfont Lemma \ref{lemma:infsup}}. Then,
 \begin{equation}\normalfont \label{eq:CF_1}
\lambda ^\uparrow _k (\textbf{A}) = \min_{S\in \mathcal{S}_k^*}\max_{{\normalfont \textbf{w}}\in S} R_{\textbf{A}}(\normalfont{\textbf{w}}),
 \end{equation}
 and
  \begin{equation}\normalfont\label{eq:CF_2}
\lambda ^\uparrow _k (\textbf{A}) =  \max_{S\in \mathcal{S}_{n-k+1}^*}\min_{{\normalfont \textbf{w}}\in S} R_{\textbf{A}}(\normalfont{\textbf{w}}).
 \end{equation}
\end{theorem}
\noindent We are now able to prove a theorem that is of particular interest to us in this paper:
\begin{theorem}\label{thm:theonetoprove}
\noindent Let $n$ be a non-zero natural integer strictly superior to one. Let \normalfont{\textbf{A}} \textcolor{black}{\textit{be a zero-trace}} $n\times n$ \textit{Hermitian matrix partitioned as}
\begin{eqnarray*}
\textbf{A} = 
 \left(
  \begin{array}{ll}
   \textbf{B}      &  \textbf{D} \\
   \textbf{D}^\dag &  \textbf{C}  \\
  \end{array} 
 \right)
\end{eqnarray*}
\textit{with}
\begin{equation*}
0\succeq \textbf{B} \in \mathbb{C}^{m\times m} \quad \mathrm{and} \quad 0\preceq \textbf{C} \in \mathbb{C}^{(n-m)\times (n-m)}
\end{equation*}
\textit{and assuming} $1 \leq m < n$. \textit{Let} $\textbf{A}_d$ \textit{be the zero-trace Hermitian matrix composed with the diagonal blocks of} \textbf{A}, i.e.,
\begin{eqnarray}\label{eq:matrix_diagonal}
\textbf{A}_d \coloneqq \textbf{B} \oplus \textbf{C}.
\end{eqnarray}
\textit{Then,}
\begin{equation*}
S_+(\textbf{A})\geq S_+(\textbf{A}_d).
\end{equation*}
\end{theorem}
\begin{proof} We use the notations of Lemma \ref{lemma:pioverpic} to denote inertia of \textbf{A}, \textbf{B}, and \textbf{C}. Let $\textbf{A}_a$ be the matrix composed with the anti-diagonal blocks of \textbf{A}, i.e.,
\begin{eqnarray}\label{eq:matrix_antidiagonal}
\textbf{A}_a \coloneqq \textbf{A} - \textbf{A}_d =
 \left(
  \begin{array}{ll}
   \textbf{0}_m      &  \textbf{D} \\
   \textbf{D}^\dag &  \textbf{0}_{(n-m)}  \\
  \end{array} 
 \right).
\end{eqnarray}
\textcolor{black}{We highlight that $\mathrm{In}(\textbf{A}_d) = \mathrm{In}(\textbf{B})+\mathrm{In}(\textbf{C)}$.} 
Let $k$ be a natural integer such that 
\begin{equation}\label{eq:kintegerPROOF1}
n\geq k \geq n-\pi_C+1-\delta_{\pi_C,0}.
\end{equation}
According to Lemma \ref{lemma:pioverpic} and according to the hypotheses of the theorem — in particular, the fact that \textbf{A} is zero-trace — for this value of $k$, we have that $\lambda_k^\uparrow(\textbf{A})$ and $\lambda_k^\uparrow(\textbf{A}_d)$ are both positive or zero. Let $\mathcal{S}_k$ be the set of every $k$--dimensional subspaces of $\mathbb{C}^n$. We define the corresponding $\mathcal{S}_k^*$ set as 
 \begin{equation*} 
 \mathcal{S}_k^* \coloneqq \bigcup _{\mathcal{S} \in \mathcal{S}_k} \left(\mathcal{S}\setminus\left\lbrace \textbf{0}_{\mathcal{S}}\right\rbrace\right).
 \end{equation*}
 We know that $n-k+1 \leq n-m$. Let $\left\lbrace \textbf{w}_{k},\ldots ,\textbf{w}_n \right\rbrace$ be a set of eigenvectors of $\textbf{C}$:
\begin{eqnarray*}
\forall j \in \llbracket k, n\rrbracket, \,  R_\textbf{C}(\textbf{w}_j) = \lambda_{j- m}^\uparrow(\textbf{C}) \geq 0.
\end{eqnarray*}
From this set, we can build another set of vectors:
\begin{eqnarray}\label{eq:qj_CF}
\forall j \in \llbracket k, n\rrbracket, \, \textbf{q}_j \coloneqq 
 \left(\!\!
  \begin{array}{l}
   \textbf{0}_{m\times 1} \\
   \textbf{w}_j   \\
  \end{array} \!\!\!
 \right).
\end{eqnarray}
Therefore,
\begin{equation}\label{eq:lambdajgeq0}
\forall j \in  \llbracket k, n\rrbracket, \, R_{\textbf{A}_d}(\textbf{q}_j) = \lambda_j^\uparrow(\textbf{A}_d) \geq 0.
\end{equation}
Let $\mathcal{Q}_k$ be defined as $\displaystyle
\mathcal{Q}_k \coloneqq \mathrm{span}(\textbf{q}_k,\ldots,\textbf{q}_n).$ According to the structure of $\textbf{A}_a$ in \eqref{eq:matrix_antidiagonal},
\begin{equation}\label{eq:Aazero}
\forall \textbf{w} \in \mathcal{Q}_k,\, \textbf{w}^\dag\textbf{A}_a\textbf{w} = 0.
\end{equation}
We define the corresponding $\mathcal{Q}_k^*$ set as $\displaystyle
 \mathcal{Q}_k^* \coloneqq   \mathcal{Q}_k \setminus\left\lbrace \textbf{0}_{n\times 1}\right\rbrace.$ Let $\mathcal{S}^*$ be any element of $\mathcal{S}_k^*$. Then,
\begin{equation*}
\dim(\mathcal{S}^*) + \dim(\mathcal{Q}_k^*) = k + (n-k+1) = n+1.
\end{equation*}
According to Lemma \ref{lemma:subspaceintersection}, and according to the definition of $\mathcal{S}^*$ and of $\mathcal{Q}^*_k$ we have that 
$$(\mathcal{S}^* \cap \mathcal{Q}^*_k  \neq \emptyset) \; \mathrm{and} \; (\forall \textbf{w}\in \mathcal{S}^* \cap \mathcal{Q}^*_k, \, \textbf{w} \neq \textbf{0}_{n\times 1}).$$ 
According to proposition \eqref{eq:subsetsupinf1} from Lemma \ref{lemma:subsetssupinf}, we find
\begin{equation}\label{eq:step1CF}
\sup_{\textbf{w} \in \mathcal{S}^*} R_{\textbf{A}}(\textbf{w}) \geq \sup _{\textbf{w} \in \mathcal{S}^*\cap \mathcal{Q}_k^*}R_{\textbf{A}}(\textbf{w}).
\end{equation}
Since $\textbf{A} = \textbf{A}_d+\textbf{A}_a$, we can use proposition \eqref{eq:Aazero} to transform \eqref{eq:step1CF} into
\begin{equation}\label{eq:step2CF}
\sup_{\textbf{w} \in \mathcal{S}^*} R_{\textbf{A}}(\textbf{w}) \geq \sup _{\textbf{w} \in \mathcal{S}^*\cap \mathcal{Q}_k^*}R_{\textbf{A}_d}(\textbf{w}).
\end{equation}
Using successively propositions \eqref{eq:subsetsupinf2} and \eqref{eq:subsetsupinf3} from Lemma \ref{lemma:subsetssupinf}, we obtain
\begin{equation*}
\sup_{\textbf{w} \in \mathcal{S}^*} R_{\textbf{A}}(\textbf{w}) \geq \inf _{\textbf{w} \in \mathcal{Q}_k^*}R_{\textbf{A}_d}(\textbf{w}).
\end{equation*}
The $\textbf{q}_k$ vector minimizes $\displaystyle
\left\lbrace R_{\textbf{A}}(\textbf{w}) \, : \, \textbf{w} \in \mathcal{Q}_k^*\right\rbrace.$ Therefore, $\displaystyle\inf _{\textbf{w} \in \mathcal{Q}_k^*}R_{\textbf{A}_d}(\textbf{w})$ is in fact $\displaystyle\min _{\textbf{w} \in \mathcal{Q}_k^*}R_{\textbf{A}_d}(\textbf{w})$:
\begin{equation}\label{eq:inequalityintCF}
\sup_{\textbf{w} \in \mathcal{S}^*} R_{\textbf{A}}(\textbf{w}) \geq \min _{\textbf{w} \in \mathcal{Q}_k^*}R_{\textbf{A}_d}(\textbf{w}) = \lambda _k^\uparrow(\textbf{A}_d).
\end{equation}
 According to \eqref{eq:pre-CF_1} in Lemma \ref{lemma:infsup}, we actually know that there is one element of $\mathcal{S}_k^*$ which minimizes $ \{\sup_{\textbf{w}\in S} R_{\textbf{A}}(\textbf{w})\, : \, S \in \mathcal{S}_k^*\}$. Denoting it $\mathcal{S}_0$, we have
\begin{equation}\label{eq:pre-end-CF}
\inf _{{S}\in\mathcal{S}_k^*} \sup_{\textbf{w} \in {S}} R_{\textbf{A}}(\textbf{w}) = \sup_{\textbf{w}\in\mathcal{S}_0}R_{\textbf{A}}(\textbf{w}).
\end{equation}
Combining \eqref{eq:pre-CF_1} in Lemma \ref{lemma:infsup} with \eqref{eq:CF_1} in Theorem \ref{theorem:courant-fischer}, we can recognize $\lambda_{k}^\uparrow(\textbf{A})$ as the left-hand side of \eqref{eq:pre-end-CF}. The inequality in \eqref{eq:inequalityintCF} is true for any $\mathcal{S}^*$ in $\mathcal{S}_k^*$. It is \textit{a fortiori} true when $\mathcal{S}^*$ is equal to $\mathcal{S}_0$. Hence, we can deduce, using \eqref{eq:lambdajgeq0}, that
\begin{equation}\label{eq:pre-result-CF}
\lambda_{k}^\uparrow(\textbf{A}) \geq   \lambda _k^\uparrow(\textbf{A}_d) \geq 0.
\end{equation}
Inequality \eqref{eq:pre-result-CF} is true for any $k$ between $n-\pi_C+1-\delta_{\pi_C,0} \eqqcolon n_C$ and $n$. Therefore,
\begin{equation}\label{eq:pre-result-CF-sum}
\sum_{k=n_C}^n \lambda_{k}^\uparrow(\textbf{A}) \geq \sum_{k=n_C}^n \lambda_{k}^\uparrow(\textbf{A}_d).
\end{equation}
The right-hand side of \eqref{eq:pre-result-CF-sum} is the sum of positive eigenvalues of $\textbf{A}_d$, i.e., $S_+( \textbf{A}_d)$. Applying Lemma \ref{lemma:S+IAleqS+A} with $I = \llbracket n_C,n\rrbracket$ and $\alpha = S_+( \textbf{A}_d)$ we finally get the desired result.
\end{proof}
\noindent An alternative proof is possible using the fact that since \textbf{A} is zero-trace in Theorem \ref{thm:theonetoprove}, $S_+(\textbf{A}) = |S_-(\textbf{A})|$. The procedure is then very similar to the above proof, but starting with $1 \leq k \leq \nu_B+\delta_{\nu_B,0}$ instead of \eqref{eq:kintegerPROOF1}, and using \eqref{eq:pre-CF_2} in Lemma \ref{lemma:infsup} and \eqref{eq:CF_2} in Theorem \ref{theorem:courant-fischer} together with Lemma \ref{lemma:S-IAgeqS-A} for $S_-^I(\textbf{A})$ rather than Lemma \ref{lemma:S+IAleqS+A}. The $\mathcal{Q}_k$ space has also to be adapted to the subspace of interest: instead of the $\textbf{q}_j$ defined in \eqref{eq:qj_CF} we would rather use 
\begin{eqnarray*}
\forall j \in \llbracket 1,k\rrbracket, \, \textbf{q}_j' \coloneqq 
 \left(\!\!
  \begin{array}{l}
   \textbf{w}_j'\\
   \textbf{0}_{(n-m)\times 1} \\
  \end{array} \!\!\!
 \right).
\end{eqnarray*}
where this time the $\textbf{w}_j'$ vectors would be eigenvectors of \textbf{B}.
\begin{proposition}\label{prop:gammarlxtracesupgamma}
 The trace of the relaxed detachment/attachment density matrices is superior or equal to the trace of the unrelaxed detachment/attachment density matrices.
\end{proposition} 
\begin{proof}
Applying Theorem \ref{thm:theonetoprove} using $\textbf{A} = \bm{\gamma}^\Delta_{rlx}$, with
\begin{eqnarray*}
\bm{\gamma}^\Delta_{rlx} = \left(\begin{array}{cc}
\textbf{C}_1 & \textbf{Z} \\ \textbf{Z}\tmat & \textbf{C}_2
\end{array}\right),
\end{eqnarray*}
 and $\textbf{A}_d = \bm{\gamma}^\Delta = \textbf{C}_1\oplus \textbf{C}_2$, where $\textbf{C}_1 = (-\textbf{XX}\tmat - \textbf{YY}\tmat) \preceq 0$ and $\textbf{C}_2 = (\textbf{X}\tmat\textbf{X}+\textbf{Y}\tmat\textbf{Y}) \succeq 0$, we finally obtain the following important inequality:
\begin{equation}\label{eq:lowerbound}
S_+(\bm{\gamma}^\Delta_{rlx}) = \vartheta^{rlx} \geq \vartheta = S_+(\bm{\gamma}^\Delta),
\end{equation}
proving that the unrelaxed detachment/attachment density integral — hence, the trace of the detachment and attachment density matrices — is a lower bound to the relaxed one.
\end{proof}
\noindent In particular, setting $\textbf{Y}$ equal to zero in $\textbf{C}_1$ and $\textbf{C}_2$ proves the conjecture of Head-Gordon and co-workers relatively to the configuration interaction singles method.


\subsection{Derivation using Cauchy's generalized interlacing theorem}
\noindent In this section we will provide an alternative proof, using Cauchy's generalized interlacing theorem, of the fact that the trace of the relaxed detachment/attachment density matrices, hence the integral over all the space of the relaxed detachment/attachment one-particle densities, is superior or equal to the unrelaxed one — i.e., Proposition \ref{prop:gammarlxtracesupgamma} — for the methods considered in this paper. The following theorem is a known result from matrix analysis, and corresponds to the ``interleaving theorem'' invoked by Head-Gordon and co-workers in their conjecture \cite{head-gordon_analysis_1995}:
\noindent \begin{theorem}[Cauchy's interlacing] \textit{Let} ${\normalfont\textbf{A}}$ \textit{be a bordered} $n\times n$ \textit{Hermitian matrix partitioned as}
$${\normalfont{\textbf{A} = \left( \begin{array}{cc}
\textbf{B} & \textbf{y} \\
\textbf{y}^\dag & a
\end{array}\right)}}$$
\textit{where} $a$ \textit{is real}, ${\normalfont{\textbf{B}}}$ belongs to $\mathbb{C}^{(n-1) \times (n-1)}$, \textit{and} ${\normalfont{\textbf{y}}}$ belongs to $\mathbb{C}^{(n-1)\times 1}$. \textit{Let} ${\normalfont{\bm{\alpha}^\downarrow = (\alpha ^\downarrow _1,\ldots , \alpha^\downarrow _n)}}$ \textit{and} ${\normalfont{\bm{\beta}^\downarrow = (\beta ^\downarrow _1,\ldots , \beta^\downarrow _{n-1})}}$ \textit{be the} $n$\textit{--tuple of the eigenvalues of} ${\normalfont{\textbf{A}}}$ \textit{sorted in the decreasing order and the} $(n-1)$\textit{--tuple of the eigenvalues of }${\normalfont{\textbf{B}}}$ \textit{sorted in decreasing order, respectively. Then,}
$${\normalfont{\alpha _1^\downarrow \geq \beta _1^\downarrow \geq \alpha _2^\downarrow \geq \cdots \geq \beta _{n-1}^\downarrow \geq \alpha _n^\downarrow .}}$$
\end{theorem}
\noindent A proof can be found in Ref. \cite{bhatia_matrix_1997}, relation (III.1). Its first natural extension, also known in the literature as ``Cauchy's generalized interlacing theorem'', is named ``Eigenvalue embedding theorem I'' below — see Theorem \ref{theo:eigembI} or its dual, i.e., Theorem \ref{theo:eigembdualI} in Appendix —, for the eigenvalues do not really ``interlace'' anymore. Theorem \ref{theo:eigembI} is one of the two twin theorems we will use here. Both can be seen as a consequence of the \textit{separation theorem} by Poincaré \cite{bhatia_matrix_1997}, i.e., in terms of compressions of Hermitian operators. For Theorem \ref{theo:eigembI} and its dual we are interested in the compression of Hermitian operators to a given subspace; for Theorem \ref{theo:eigembII} and its dual — i.e., Theorem \ref{theo:eigembdualII} in Appendix — we are interested in the compression of the same Hermitian operator to the complementary space relatively to Theorem \ref{theo:eigembI}. The complementarity of the two theorems will then be used for deriving our first boundary value for the relaxed detachment/attachment trace. Notice that since ``Cauchy's generalized interlacing theorem'' is most often introduced and proved in the shape of \ref{theo:eigembI}, we provide here a proof for the second compression, i.e., for theorem \ref{theo:eigembII} instead.

The $\hat{\gamma}^\Delta$ map and the $\hat{\gamma}^\Delta_{rlx}$ map, i.e., our one-body reduced unrelaxed and relaxed difference density operators \cite{etienne_towards_2021}, living in the linear span of the $L$-dimensional
$$ \mathcal{B} = (\mathcal{B}_o,\mathcal{B}_v)$$
ordered basis of one-particle-state functions, respectively have the $\bm{\gamma}^\Delta$ and the $\bm{\gamma}^\Delta_{rlx}$ matrix representations in $\mathcal{B}$. Here, $\mathcal{B}_o \coloneqq \left(\varphi_i\right)_{i\in \llbracket 1, N \rrbracket}$, and $\mathcal{B}_v \coloneqq \left(\varphi_i\right)_{i\in \llbracket N+1,L\rrbracket}$. If we denote by $\mathcal{F}$, $\mathcal{F}_o$, and $\mathcal{F}_v$ respectively the real linear span of $\mathcal{B}$, $\mathcal{B}_o$, and $\mathcal{B}_v$, we have that $\hat{\gamma}^\Delta$ and $\hat{\gamma}^\Delta_{rlx}$ are two $\mathcal{F} \longrightarrow \mathcal{F}$ maps, and that the compression of these two operators to $\mathcal{F}_o$ using
\begin{align*}
\hat{R}_o \;:\; \left(\mathcal{F} \longrightarrow \mathcal{F}\right) &\longrightarrow \left(\mathcal{F}_o \longrightarrow \mathcal{F}_o\right) \\
\hat{A} &\longmapsto \hat{C}_o^\dag \hat{A}\hat{C}_o^{\textcolor{white}{\dag}},
\end{align*}
leads to a map that has identical matrix representation (i.e., $\textbf{C}_1$ in \eqref{eq:DeltaAB}) in $\mathcal{B}_o$ when $\hat{A}$ is either $\hat{\gamma}^\Delta$ or $\hat{\gamma}^\Delta_{rlx}$ if the compression operator $\hat{C}_o$ is the $\mathcal{F}_o\longrightarrow \mathcal{F}$ injection map, i.e., if it has the rectangular
$$ \mathcal{M}\left(\hat{C}_o, \mathcal{B}_o, \mathcal{B}\right) = \left(\begin{array}{c}
\textbf{I}_{o} \\ \textbf{0}_{vo}
\end{array} \right)$$
 matrix representation, where $\textbf{I}_o$ is the $N\times N$ identity matrix, and $\textbf{0}_{vo}$ is the $(L-N)\times N$ zero matrix. The following theorem corresponds to this first compression — a proof can be found in section 4.3 (theorem 4.3.28) of Ref. \cite{horn_matrix_2012}.
\noindent \begin{theorem}[Eigenvalue embedding I]\label{theo:eigembI} \textit{Let} ${\normalfont\textbf{A}}$ \textit{be an} $n\times n$ \textit{Hermitian matrix partitioned as}
$${\normalfont{\textbf{A} = \left( \begin{array}{cc}
\textbf{B}^{\textcolor{white}{*}} & \textbf{D} \\
\textbf{D}^\dag & \textbf{C}
\end{array}\right)}}$$
\textit{with} ${\normalfont{\textbf{B}}} \in \mathbb{C}^{m \times m},$ \textit{assuming that} $(m < n)$. \textit{Let} ${\normalfont{\bm{\alpha}^\uparrow = (\alpha ^\uparrow _1,\ldots , \alpha^\uparrow _n)}}$ \textit{and} ${\normalfont{\bm{\beta}^\uparrow = (\beta ^\uparrow _1,\ldots , \beta^\uparrow _m)}}$ \textit{be the} $n$\textit{--tuple of the eigenvalues of} ${\normalfont{\textbf{A}}}$ \textit{sorted in the increasing order and the} $m$\textit{--tuple of the eigenvalues of }${\normalfont{\textbf{B}}}$ \textit{sorted in increasing order, respectively. Then,}
$$ \forall i \in \llbracket 1, m \rrbracket, \, \alpha _i^\uparrow \leq \beta _i ^\uparrow \leq \alpha _{i+n-m}^\uparrow.$$
\end{theorem}
\noindent The compression of $\hat{\gamma}^\Delta$ and $\hat{\gamma}^\Delta_{rlx}$ to $\mathcal{F}_v$ using
\begin{align*}
\hat{R}_v \;:\; \left(\mathcal{F}\longrightarrow \mathcal{F}\right) &\longrightarrow \left(\mathcal{F}_v\longrightarrow\mathcal{F}_v\right) \\
\hat{A} &\longmapsto \hat{C}_v^\dag \hat{A}\hat{C}_v^{\textcolor{white}{\dag}},
\end{align*}
leads to a map that has identical matrix representation (i.e., $\textbf{C}_2$ in \eqref{eq:DeltaAB}) in $\mathcal{B}_v$ when $\hat{A}$ is either $\hat{\gamma}^\Delta$ or $\hat{\gamma}^\Delta_{rlx}$ if the compression operator $\hat{C}_v$ is the $\mathcal{F}_v\longrightarrow \mathcal{F}$ injection map, i.e., if it has the rectangular
$$ \mathcal{M}\left(\hat{C}_v, \mathcal{B}_v, \mathcal{B}\right) = \left(\begin{array}{c}
\textbf{0}_{ov} \\ \textbf{I}_{v}
\end{array} \right)$$
 matrix representation, where $\textbf{I}_v$ is the $(L-N)\times (L-N)$ identity matrix, and $\textbf{0}_{ov}$ is the $N \times (L-N)$ zero matrix. The following theorem corresponds to this second compression, and is a twin theorem to Theorem \ref{theo:eigembI}:
\noindent \begin{theorem}[Eigenvalue embedding II]\label{theo:eigembII} \textit{Let} ${\normalfont\textbf{A}}$ \textit{be an} $n\times n$ \textit{Hermitian matrix partitioned as }
$${\normalfont\textbf{A}} = \left( \begin{array}{cc}
{\normalfont\textbf{B}}^{\textcolor{white}{*}} & {\normalfont\textbf{D}} \\
{\normalfont\textbf{D}}^\dag & {\normalfont\textbf{C}}
\end{array}\right)
$$
\textit{with} ${\normalfont\textbf{C}} \in \mathbb{C}^{(n-m) \times (n-m)},$ \textit{assuming that} $(m < n)$. \textit{Let} $\bm{\alpha}^\downarrow = (\alpha ^\downarrow _1,\ldots , \alpha^\downarrow _n)$ \textit{and} $\bm{\gamma}^\downarrow = (\gamma ^\downarrow _1,\ldots , \gamma^\downarrow _{n-m})$ \textit{be the} $n$\textit{--tuple of the eigenvalues of} ${\normalfont\textbf{A}}$ \textit{sorted in the decreasing order and the} $(n-m)$\textit{--tuple of the eigenvalues of }${\normalfont\textbf{C}}$ \textit{sorted in decreasing order, respectively. Then,}
$$\normalfont\forall i \in \llbracket 1, (n-m) \rrbracket, \, \alpha _i^\downarrow \geq \gamma _i ^\downarrow \geq \alpha _{i+m}^\downarrow.$$
\end{theorem}  
\begin{proof}
The $\textbf{A}$ matrix can be seen as the $\mathcal{A} \longrightarrow \mathcal{A}$ linear map $\hat{A}$, with $\mathcal{A} \coloneqq \mathrm{span}\,\mathcal{B}_c$, in which $\mathcal{B}_c$ is the orthonormal ordered basis $(\textbf{c}_i)_{i\in\llbracket 1,n\rrbracket}$ such that
$$\forall i \in \llbracket 1, n\rrbracket, \, \hat{A}\textbf{c}_i = \sum_{j=1}^n (\textbf{A})_{j,i}\textbf{c}_i.$$
$\mathcal{A}$ can be seen as $\mathcal{C}_1\oplus \mathcal{C}_2$, where $\mathcal{C}_1 = \mathrm{span}(\textbf{c}_i)_{i\in\llbracket 1,m\rrbracket}$, and $\mathcal{C}_2 = \mathrm{span}(\textbf{c}_i)_{i\in\llbracket m+1,n\rrbracket}$. We also define $\mathcal{C}^*_2 \coloneqq \mathcal{C}_2\setminus \{\textbf{0}_{\mathcal{C}_2}\}$. Let $\hat{C}_2$ be the $\mathcal{C}_2\longrightarrow\mathcal{A}$ injection map:
\begin{align*}
\forall \textbf{c}\in \mathcal{C}_2,\, \textbf{c}&=\sum_{j=m+1}^nc_j\textbf{c}_j,\\
\hat{C}_2\textbf{c}&=\sum_{i=1}^m 0\textbf{c}_i + \sum_{j=m+1}^nc_j\textbf{c}_j.
\end{align*}
The $\hat{C} = \hat{C}_2^\dag\hat{A}\hat{C}_2$ map is the compression of $\hat{A}$ on $\mathcal{C}_2$:
\begin{align*}
\hat{C} \, : \ \mathcal{C}_2 &\longrightarrow \mathcal{C}_2 \\
\textbf{w}&\longmapsto \hat{C}\textbf{w} = \textbf{C}\textbf{w}.
\end{align*}
We have that, using again Definition \ref{def:rayleigh} for $R$,
\begin{equation}\label{eq:inCinA}
\forall \textbf{w}\in \mathcal{C}_2^*,\, R_\textbf{A}(\hat{C}_2\textbf{w})=R_\textbf{C}(\textbf{w}).
\end{equation}
Let $j$ be any natural integer between 1 and $(n-m)$. Let $\textbf{U}^\downarrow \coloneqq (\textbf{u}^\downarrow_i)_{i\in\llbracket 1,(n-m)\rrbracket}$ be the ordered basis of eigenvectors of $\hat{C}$ corresponding to the eigenvalues of $\hat{C}$ ordered in decreasing order $(\lambda_1^\downarrow (\textbf{C}) \geq \cdots \geq \lambda_{n-m}^\downarrow (\textbf{C}))$. We set a $j$--tuple, $\textbf{U}_j^\downarrow \coloneqq (\textbf{u}_i^\downarrow)_{i\in\llbracket 1,j\rrbracket}^{\textcolor{white}{\downarrow}}$, and build
$$\mathcal{C}_{2,j} \coloneqq \mathrm{span}(\textbf{U}^\downarrow_j) \quad \mathrm{and} \quad \mathcal{C}_{2,j}^* \coloneqq \mathcal{C}_{2,j} \setminus \{\textbf{0}_{\mathcal{C}_{2,j}}\}.$$
Let $\textbf{V}^\downarrow \coloneqq (\textbf{v}^\downarrow_i)_{i\in\llbracket 1,n\rrbracket}$ be the ordered basis of eigenvectors of $\hat{A}$ corresponding to the eigenvalues of $\hat{A}$ ordered in decreasing order $(\lambda_1^\downarrow (\textbf{A}) \geq \cdots \geq \lambda_n^\downarrow (\textbf{A}))$. We set an $(n-j+1)$--tuple, $\textbf{V}_j^\downarrow \coloneqq (\textbf{v}_i^\downarrow)_{i\in\llbracket j,n\rrbracket}^{\textcolor{white}{\downarrow}}$, and build
$$\mathcal{A}_{j} \coloneqq \mathrm{span}(\textbf{V}^\downarrow_j) \quad \mathrm{and} \quad \mathcal{A}_{j}^* \coloneqq \mathcal{A}_{j} \setminus \{\textbf{0}_{\mathcal{A}_{j}}\}\subset \mathbb{C}^n.$$
We write any element of $\mathcal{C}_{2,j}^*$, say $\textbf{u}$, as $\textbf{u} = \sum_{i=1}^ju_i\textbf{u}^\downarrow_i$, we denote by $\vartheta_\textbf{u}$ the $\textbf{u}^\dag\textbf{u}$ strictly positive number, and see that
\begin{align*}
R_\textbf{C}(\textbf{u}) = \vartheta_\textbf{u}^{-1} \sum_{i=1}^j|u_i|^2\lambda^\downarrow_i(\textbf{C})\geq \vartheta_\textbf{u}^{-1} \sum_{i=1}^j|u_i|^2\lambda^\downarrow_j(\textbf{C}).
\end{align*}
Hence, for any $\textbf{u}$ in $\mathcal{C}_{2,j}^*$ we have $R_\textbf{C}(\textbf{u}) \geq \lambda_j^\downarrow (\textbf{C})$. We conclude that
\begin{equation}\label{eq:minuC2eqlambdaC}
(\forall \textbf{u} \in \mathcal{C}_{2,j}^*,\, R_\textbf{C}(\textbf{u}) \geq \lambda_j^\downarrow (\textbf{C})) \land (\textbf{u}_j\in \mathcal{C}_{2,j}^*) \Longrightarrow \min_{\textbf{u}\in \mathcal{C}_{2,j}^*} R_{\textbf{C}}(\textbf{u}) = \lambda_{j}^\downarrow (\textbf{C}).
\end{equation}
We write any element of $\mathcal{A}_{j}^*$, say $\textbf{v}$, as $\textbf{v} = \sum_{i=j}^n v_i\textbf{v}^\downarrow_i$, we denote by $\vartheta_\textbf{v}$ the $\textbf{v}^\dag\textbf{v}$ strictly positive number, and see that
\begin{align*}
R_\textbf{A}(\textbf{v}) = \vartheta_\textbf{v}^{-1} \sum_{i=j}^n|v_i|^2\lambda^\downarrow_i(\textbf{A})\leq \vartheta_\textbf{v}^{-1} \sum_{i=j}^n|v_i|^2\lambda^\downarrow_j(\textbf{A}).
\end{align*}
Hence, we conclude that $\lambda_j^\downarrow (\textbf{A})$ constitutes a certain boundary value:
\begin{equation}\label{eq:upperboundjLambdaA}
\forall \textbf{v} \in \mathcal{A}_{j}^*,\, R_\textbf{A}(\textbf{v}) \leq \lambda_j^\downarrow (\textbf{A}).
\end{equation}
We now define the set of the images to the $\mathcal{C}_{2,j}^* \longrightarrow \mathcal{A}$ injection map and notice that it is a subspace of $\mathbb{C}^n$ — as was $\mathcal{A}_j ^*$ above —, which will be useful later:
$$\mathcal{C}_{2,j}^{*,\perp} \coloneqq \{\hat{C}_2\textbf{u} \, : \, \textbf{u} \in \mathcal{C}_{2,j}^*\} \subset \mathbb{C}^n,$$ 
and immediately deduce the identity between two minima — one related to $\mathcal{C}_{2,j}^*$, and the other related to $\mathcal{C}_{2,j}^{*,\perp}$:
\begin{equation}\label{eq:minC2eqminC2perp}
\min_{\textbf{u}\in\mathcal{C}_{2,j}^*}R_\textbf{A}(\hat{C}_2\textbf{u}) = \min_{\textbf{u}'\in\mathcal{C}_{2,j}^{*,\perp}}R_\textbf{A}(\textbf{u}') \eqqcolon {A}_j.
\end{equation}
We also know that, by definition of the minimum that has just been written above,
\begin{equation}\label{eq:RvAgeqRvAmin}
\forall \textbf{u}' \in \mathcal{C}_{2,j}^{*,\perp},\, R_\textbf{A}(\textbf{u}') \geq {A}_j.
\end{equation}
We then have, according to \eqref{eq:inCinA}, \eqref{eq:minuC2eqlambdaC}, \eqref{eq:minC2eqminC2perp} and \eqref{eq:RvAgeqRvAmin}, that
$$\lambda^\downarrow _j (\textbf{C}) = \min_{\textbf{u}\in\mathcal{C}_{2,j}^*}R_\textbf{C}(\textbf{u}) = \min_{\textbf{u}\in\mathcal{C}_{2,j}^*}R_\textbf{A}(\hat{C}_2\textbf{u}),$$
i.e., $\lambda^\downarrow _j (\textbf{C})$ is a boundary value for the values of $R_\textbf{A}$ when its argument lies in $\mathcal{C}_{2,j}^{*,\perp}$:
\begin{equation}\label{eq:foralluprime}
\forall \textbf{u}' \in \mathcal{C}_{2,j}^{*,\perp},\, \lambda^\downarrow _j (\textbf{C}) \leq R_\textbf{A}(\textbf{u}').
\end{equation}
The $(n-j+1)$--dimensional $\mathcal{A}^*_{j}$ space and the $j$--dimensional $\mathcal{C}_{2,j}^{*,\perp}$ space are two subspaces of $\mathbb{C}^n$. According to Lemma \ref{lemma:subspaceintersection},
$$ (\mathcal{A}^*_j\subset \mathbb{C}^n \land \mathcal{C}_{2,j}^{*,\perp} \subset \mathbb{C}^n) \land(\mathrm{dim}(\mathcal{A}^*_j) + \mathrm{dim}(\mathcal{C}_{2,j}^{*,\perp})> n) \Longrightarrow \exists \textbf{p}\in \mathcal{A}^*_{j} \cap \mathcal{C}_{2,j}^{*,\perp}, \, \textbf{p}\neq \textbf{0}_{n\times 1}.$$
A vector in $\mathcal{A}^*_{j} \cap \mathcal{C}_{2,j}^{*,\perp}\setminus \{\textbf{0}_{n\times 1}\}$ will simultaneously satisfy \eqref{eq:upperboundjLambdaA} and \eqref{eq:foralluprime}:
$$\forall \textbf{p} \in \mathcal{A}^*_{j} \cap \mathcal{C}_{2,j}^{*,\perp}\setminus \{\textbf{0}_{n\times 1}\}, \, \lambda_j^\downarrow (\textbf{C}) \leq R_\textbf{A}(\textbf{p}) \leq \lambda_j^\downarrow (\textbf{A}),$$
which is possible only if $\lambda_j^\downarrow (\textbf{C})  \leq \lambda_j^\downarrow (\textbf{A})$. This result is valid for any $j$ in $\llbracket 1, (n-m)\rrbracket$:
\begin{equation}\label{eq:eigembIIpart1}
\forall j \in \llbracket 1,(n-m)\rrbracket, \,   \lambda_j^\downarrow (\textbf{A}) \geq \lambda_j^\downarrow (\textbf{C}).
\end{equation}
Applying \eqref{eq:eigembIIpart1} to $-\textbf{A}$ instead of $\textbf{A}$ gives
$$\forall j \in \llbracket 1, (n-m)\rrbracket, \,    \lambda_j^\downarrow (-\textbf{A}) \geq \lambda_j^\downarrow (-\textbf{C}),$$
i.e.,
\begin{equation}\label{eq:minslambda}
\forall j\in \llbracket 1, (n-m)\rrbracket,\, -\lambda^\downarrow _{n-j+1}(\textbf{A}) \geq -\lambda_{(n-m)-j+1}^\downarrow (\textbf{C}).
\end{equation}
For every $j$ in $\llbracket 1, (n-m)\rrbracket$, setting $j_p \coloneqq \left[(n-m)-j+1\right] \in \llbracket 1,(n-m)\rrbracket$ turns \eqref{eq:minslambda} into
\begin{equation} \label{eq:IIpre-conclusion}
\forall j_p\in \llbracket 1, (n-m)\rrbracket,\, \lambda^\downarrow _{j_p+m}(\textbf{A}) \leq \lambda_{j_p}^\downarrow (\textbf{C}).
\end{equation}
Combining \eqref{eq:eigembIIpart1} with \eqref{eq:IIpre-conclusion} completes the proof.
\end{proof}
\noindent An alternative, shorter but less instructive proof of Theorem \ref{theo:eigembII} is provided in Appendix \ref{app:eigembII}. For the sake of completeness, we introduce and prove the dual of theorems \ref{theo:eigembI} and \ref{theo:eigembII} in Appendix \ref{app:dualeigembI} and \ref{app:dualeigembII}, respectively. 

We are now equipped for proving that Theorem \ref{thm:theonetoprove} is in fact a corollary of Theorem \ref{theo:eigembI} and Theorem \ref{theo:eigembII}:
\begin{proof}
We are in the conditions of Theorem \ref{thm:theonetoprove}. Let $(\lambda^\downarrow_i(\textbf{A}))_{i \in \llbracket 1,n \rrbracket}$ be the $n$-tuple of eigenvalues of $\textbf{A}$ sorted in decreasing order. Let $(\lambda^\downarrow_i(\textbf{C}))_{i \in \llbracket 1,(n-m) \rrbracket}$ be the $(n-m)$--tuple of eigenvalues of $\textbf{C}$ sorted in decreasing order, and let $(\lambda^\downarrow_i(\textbf{B}\oplus \textbf{C}))_{i \in \llbracket 1,n \rrbracket}$ be the $n$-tuple of eigenvalues of $\textbf{B}\oplus \textbf{C}$ sorted in decreasing order. According to Theorem \ref{theo:eigembII}, we have that
\begin{equation}\label{eq:nmmfirstgeq0}
\left(\forall p \in \llbracket 1,(n-m)\rrbracket, \, \lambda_p^\downarrow(\textbf{C}) \geq 0\right) \, \Longrightarrow \, \left( \forall p \in \llbracket 1,(n-m)\rrbracket, \, \lambda_p^\downarrow (\textbf{A}) \geq \lambda_p^\downarrow(\textbf{C}) \geq 0\right).
\end{equation}
The left-hand side proposition is true due to the hypotheses — i.e., that $\textbf{C} \succeq 0$ —, hence the right-hand side proposition is true in virtue of Theorem \ref{theo:eigembII}. Since
\begin{equation*}
\forall p \in \llbracket 1,m\rrbracket,\, \lambda^\downarrow _{n-m+p}(\textbf{A}) = \lambda^\uparrow_{m-p+1}(\textbf{A}),
\end{equation*}
we find that, according to the hypotheses — i.e., that $\textbf{B} \preceq 0$ — and according to Theorem \ref{theo:eigembI},
\begin{equation}\label{eq:mlastleq0}
\forall p \in \llbracket n-m+1, n\rrbracket,\, \lambda^\downarrow_p(\textbf{A}) \leq 0.
\end{equation}
The fact that $\textbf{B} \preceq 0$ and $\textbf{C} \succeq 0$, together with \eqref{eq:nmmfirstgeq0}, gives
\begin{equation}\label{eq:pre-combination1}
 \sum _{p=1}^{n-m} \lambda^\downarrow _p(\textbf{A}) \geq \sum _{p=1}^{n-m} \lambda^\downarrow _p (\textbf{C}) =  \sum _{p=1}^{n-m} \lambda^\downarrow _p (\textbf{B}\oplus\textbf{C})  \geq 0.
\end{equation}
On the other hand, the fact that $\textbf{B} \preceq 0$ and $\textbf{C} \succeq 0$, together with \eqref{eq:mlastleq0}, gives
\begin{equation}\label{eq:pre-combination2}
 \sum _{p=n-m+1}^{n} \lambda^\downarrow _p(\textbf{A}) \leq 0 \quad \mathrm{and} \quad  \sum _{p=n-m+1}^{n} \lambda^\downarrow _p (\textbf{B}\oplus\textbf{C}) = \sum _{p=1}^{m} \lambda^\downarrow _p (\textbf{B}) \leq 0.
\end{equation}
Results in \eqref{eq:pre-combination1} and \eqref{eq:pre-combination2} can be combined to derive the following inequality:
\begin{equation*}
\sum _{p=1}^n \mathrm{max}(\lambda^\downarrow _p(\textbf{A}),0) \geq  \sum _{p=1}^{n} \mathrm{max}(\lambda^\downarrow _p(\textbf{B}\oplus\textbf{C}),0),
\end{equation*}
i.e.,
\begin{equation*}
S_+(\textbf{A})  \geq   S_+({\textbf{B}\oplus\textbf{C}}),
\end{equation*}
which completes the proof. 
\end{proof}
\noindent Since the proof of Proposition \ref{prop:gammarlxtracesupgamma} relies exclusively on the application of Theorem \ref{thm:theonetoprove}, and since we have just shown that Theorem \ref{thm:theonetoprove} can be proved using the twin theorems \ref{theo:eigembI} and \ref{theo:eigembII}, we see that the fact that $\vartheta$ is a lower boundary value to the trace of the relaxed detachment and attachment density matrices can be proved using the twin theorems \ref{theo:eigembI} and \ref{theo:eigembII}.
\begin{corollary}\label{lemma:pioverpicTWO}
In the conditions of {\normalfont Lemma \ref{lemma:pioverpic}}, we have
\begin{equation*}
(\nu_B\leq \nu \leq m)\wedge (\pi_C \leq \pi \leq (n-m)).
\end{equation*}
\end{corollary}
\begin{proof}
$(\nu_B\leq \nu)$ and $(\pi_C \leq \pi)$ were already part of Lemma \ref{lemma:pioverpic}. Combining \eqref{eq:nmmfirstgeq0} and \eqref{eq:mlastleq0} we find the two other inequalities — that should be compared to the result in Lemma \ref{lemma:piinftonminusm} where the constraints on the structure and nature of $\textbf{A}$ were less strong.
\end{proof}
\noindent In Corollary \ref{lemma:pioverpicTWO}, $\nu_B = \nu(\textbf{B}\oplus\textbf{C})$ and $\pi_C = \pi(\textbf{B}\oplus\textbf{C})$. We see that applying Corollary \ref{lemma:pioverpicTWO} using $\bm{\gamma}^\Delta_{rlx}$ as $\textbf{A}$ and $\bm{\gamma}^\Delta$ as $\textbf{B}\oplus \textbf{C}$, we find, using Definition \ref{def:orderrelationship},
$$ (N,(L-N))\geq (\pi(\bm{\gamma}^\Delta_{rlx}), \nu(\bm{\gamma}^\Delta_{rlx})) \geq (\pi(\bm{\gamma}^\Delta),\nu(\bm{\gamma}^\Delta)),$$
which should be compared with the results, derived in Ref. \cite{etienne_auxiliary_2025}, for the unrelaxed $\bm{\gamma}^\Delta$:
\begin{align*}
(|\pi_x-\pi_y| \leq \pi(\bm{\gamma}^\Delta) \leq \pi_x + \pi_y) \; &\mathrm{and} \; \pi(\bm{\gamma}^\Delta) \leq \mathrm{min}(2N,(L-N)),\\
(|\nu_x-\nu_y| \leq \nu(\bm{\gamma}^\Delta) \leq \nu_x + \nu_y) \; &\mathrm{and} \; \nu(\bm{\gamma}^\Delta) \leq \mathrm{min}(N,2(L-N)),
\end{align*}
where $\pi_x \coloneqq \pi(\textbf{X}\tmat\textbf{X})$, $\pi_y \coloneqq \pi(\textbf{Y}\tmat\textbf{Y})$, $\nu_x \coloneqq \nu(-\textbf{X}\tmat\textbf{X}) = \pi_x$, $\nu_y \coloneqq \nu(-\textbf{Y}\tmat\textbf{Y}) = \pi_y$. When both $\textbf{Z}$ and $\textbf{Y}$ are the $N\times (L-N)$ zero matrices, we have that $\bm{\gamma}^\Delta$ reduces to $(-\textbf{XX}\tmat)\oplus\textbf{X}\tmat\textbf{X}$, and
$$\pi((-\textbf{XX}\tmat)\oplus\textbf{X}\tmat\textbf{X}) = \nu((-\textbf{XX}\tmat)\oplus\textbf{X}\tmat\textbf{X}) \leq \mathrm{min}(N,(L-N)).$$


\section{Upper bound to the relaxed detachment/attachment trace}
\noindent In this section we will first prove that the sum of the positive eigenvalues of $\bm{\gamma}^\mathrm{Z}$ is equal to the sum of the singular values of $\textbf{Z}$, and we will combine this result with an important result from linear algebra to derive the upper boundary value to the charge transferred during a molecular electronic transition.
\begin{theorem}\label{prop:Z1}
Let $m$ and $n$ be two natural integers such that $1\leq m < n$. Let {\normalfont{\textbf{D}}} be an $m\times (n-m)$ complex matrix. Let {\normalfont{$\textbf{A}_a$}} be the $n\times n$ Hermitian matrix partitioned as
\begin{eqnarray*}{\normalfont{
\textbf{A}_a \coloneqq 
 \left(
  \begin{array}{ll}
   \textbf{0}_m      &  \textbf{D} \\
   \textbf{D}^\dag &  \textbf{0}_{(n-m)}  \\
  \end{array} 
 \right).}}
\end{eqnarray*}
Then, {\normalfont{$S_+(\textbf{A}_a)$}} is equal to the sum of the singular values of {\normalfont{$\textbf{D}$}}.
\end{theorem}
\begin{proof}
The singular value decomposition of the $\textbf{D}$ matrix from $\textbf{A}_a$ reads
\begin{equation*}
\exists (\textbf{V},\textbf{S},\textbf{U}) \in \mathbb{C}^{m\times m}\times \mathbb{R}_+^{m\times(n-m)}\times \mathbb{C}^{(n-m)\times(n-m)},\,\textbf{D} = \textbf{VSU}^\dag.
\end{equation*}
The matrix containing the singular values, $\textbf{S}$, is diagonal, i.e.,
$$\forall (i,j) \in \llbracket 1,m\rrbracket \times \llbracket 1,(n-m)\rrbracket, \, (i\neq j) \Longrightarrow \left(\textbf{S}\right)_{i,j} = 0.$$
We set $q \coloneqq \min(m,(n-m))$, and
\begin{equation*}
\forall p \in \llbracket1,q\rrbracket, \, s_p \coloneqq \left(\textbf{S}\right)_{p,p}.
\end{equation*}
We also have that $\textbf{V}^\dag=\textbf{V}^{-1}$ and $\textbf{U}^\dag = \textbf{U}^{-1}$. The square of $\textbf{A}_a$ can also be written
\begin{equation*}
\textbf{A}_a^\dag\textbf{A}_a = \textbf{DD}^\dag \oplus \textbf{D}^\dag\textbf{D}
\end{equation*}
and commutes with $\textbf{A}_a$. One possible eigenvalue decomposition of $\textbf{A}_a^{\dag}\textbf{A}_a$ reads
\begin{equation*}
\textbf{G}^\dag \left(\textbf{DD}^\dag \oplus \textbf{D}^\dag\textbf{D}\right)\textbf{G} = \textbf{SS}^\dag\oplus \textbf{S}^\dag\textbf{S}
\end{equation*}
with $\textbf{G} = \textbf{V}\oplus\textbf{U}$, so the squared Frobenius norm of $\textbf{D}$ is simply
\begin{equation}\label{eq:squarefrobeniusD}
\mathrm{tr}(\textbf{D}^\dag\textbf{D}) = \sum _{p=1}^{q} s_p^2.
\end{equation}
Each non-zero eigenvalue in $\textbf{SS}^\dag\oplus \textbf{S}^\dag\textbf{S}$ is degenerate with a multiplicity of two. Since the $\textbf{DD}^\dag \oplus \textbf{D}^\dag\textbf{D}$ matrix commutes with $\textbf{A}_a$, it is possible to build an orthonormal basis with vectors that are eigenvectors of both matrices. To construct this basis, we first reconsider $\textbf{V}$ as an orthonormal $m$-tuple of vectors, i.e., $\textbf{V} := \left(\textbf{v}_i\right)_{i\in \llbracket 1,m\rrbracket}$. We also reconsider $\textbf{U}$ as an orthonormal $(n-m)$-tuple of vectors, i.e., $\textbf{U} := \left(\textbf{u}_i\right)_{i\in \llbracket 1,(n-m)\rrbracket}$, so we can finally rewrite $\textbf{G}$ as an $n$-tuple of vectors, i.e., $\textbf{G} := \left(\textbf{v}_1',\ldots,\textbf{v}_m',\textbf{u}_1',\ldots,\textbf{u}_{n-m}'\right)$, with
\begin{equation*}
\forall p \in \llbracket 1, m \rrbracket, \, \textbf{v}_p' := \begin{pmatrix}
\textbf{v}_p \\ \textbf{0}_{(n-m)\times 1}
\end{pmatrix},
\end{equation*}
and
\begin{equation*}
\forall p \in \llbracket 1, (n-m)\rrbracket, \textbf{u}_p' := \begin{pmatrix}
\textbf{0}_{m\times 1} \\ \textbf{u}_p
\end{pmatrix}.
\end{equation*}
The vectors in $\textbf{V}$ and $\textbf{U}$ are ordered so that
\begin{align*}
\forall p \in \llbracket 1, q\rrbracket, \quad \textbf{D}\textbf{u}_p &= s_p \textbf{v}_p, \\
\textbf{D}^\dag \textbf{v}_p &= s_p \textbf{u}_p.
\end{align*}
We now construct $
\textbf{W} \coloneqq \left(\textbf{w}_1^+,\ldots,\textbf{w}_q^+,\textbf{w}_1^-,\ldots, \textbf{w}_q^-,\textbf{W}'\right)$, with
\begin{align*}
\forall p\in \llbracket 1, q\rrbracket, \, (\sqrt{2}\textbf{w}_p^+ &= {\textbf{v}_p' + \textbf{u}_p'}) \quad \mathrm{and} \quad
(\sqrt{2}\textbf{w}_p^- = {\textbf{v}_p' - \textbf{u}_p'}{}),
\end{align*}
and, for the last $n-2q$ columns, $
\textbf{W}' := \left(\textbf{w}'_1,\ldots,\textbf{w}_{n-2q}'\right)$ with, for every $p$ in $\llbracket 1,n-2q\rrbracket$,
\begin{equation*}
\textbf{w}_p' \in \left\lbrace \textbf{w}\in \mathbb{C}^{n\times 1} \, : \, \textbf{w} \in \mathrm{ker}\left(\textbf{A}_a\right)\;\land\; \textbf{w}^\dag\textbf{w} = 1 \right\rbrace.
\end{equation*}
We can see that the first $2q$ vectors in $\textbf{W}$ are eigenvectors of both $\textbf{A}_a$ and of its second power. We define three matrices: $\textbf{S}_2 \coloneqq \mathrm{diag}\left(s_1^2,\ldots,s_q^2\right)$,  $\textbf{S}_+ \coloneqq \mathrm{diag}\left(s_1,\ldots,s_q\right)$, and $\textbf{S}_- \coloneqq - \textbf{S}_+$. We now need to study two cases:

$\;$

\noindent \textit{Case 1.} $(n-m \neq m)$ — We have 
\begin{align}\nonumber
\textbf{W}^\dag \textbf{A}_a^{\dag}&\textbf{A}_a\textbf{W} = \textbf{S}_2\,\oplus \textbf{S}_2\, \oplus \textbf{0}_{n-2q}, \\
\textbf{W}^\dag &\textbf{A}_a\textbf{W} = \textbf{S}_+ \oplus \textbf{S}_- \oplus \textbf{0}_{n-2q}.\label{eq:eigdecD1}
\end{align}
\noindent \textit{Case 2.} $(n-m=m)$ — We have 
\begin{align} \nonumber
\textbf{W}^\dag \textbf{A}_a^{\dag}&\textbf{A}_a\textbf{W} = \textbf{S}_2\,\oplus \textbf{S}_2, \\
\textbf{W}^\dag &\textbf{A}_a\textbf{W} = \textbf{S}_+ \oplus \textbf{S}_-.\label{eq:eigdecD2}
\end{align}
We see that in both cases the sum of the positive eigenvalues of $\textbf{A}_a$ reduces to the sum of the singular values of $\textbf{A}_a$, i.e.,
\begin{equation*}
S_+(\textbf{A}_a) = \mathrm{tr}(\textbf{S}_+) = \sum_{p=1}^q s_p.
\end{equation*}
\end{proof}
\noindent Applying Theorem \ref{prop:Z1} to the case of $\bm{\gamma}^\mathrm{Z}$ from \eqref{eq:gammaZ}, i.e., with $\textbf{D}=\textbf{Z}$, gives the following results: In \textit{Case 1} as well as in \textit{Case 2}, we see that the singular value decomposition of the \textbf{Z} matrix, together with the kernel of $\bm{\gamma}^\mathrm{Z}$, fully determine the detachment and attachment density matrices related to the relaxation process: in \textit{Case 1} we have, from \eqref{eq:eigdecD1},
\begin{align*}
\bm{\gamma}^d_z &= \textbf{W}\left(\textbf{0}_{q} \oplus \textbf{S}_+ \oplus \textbf{0}_{n-2q}\right)\textbf{W}\tmat, \\
\bm{\gamma}^a_z &= \textbf{W}\left(\textbf{S}_+ \oplus \textbf{0}_{q} \oplus \textbf{0}_{n-2q}\right)\textbf{W}\tmat,
\end{align*}
while in \textit{Case 2} we have, from \eqref{eq:eigdecD2},
\begin{align*}
\bm{\gamma}^d_z &= \textbf{W}\left(\textbf{0}_{q} \oplus \textbf{S}_+ \right)\textbf{W}\tmat, \\
\bm{\gamma}^a_z &= \textbf{W}\left(\textbf{S}_+ \oplus \textbf{0}_{q} \right)\textbf{W}\tmat.
\end{align*}
Therefore, the integral of the Z-vector-related detachment/attachment densities is then
\begin{equation*}
\vartheta ^\mathrm{Z} = \sum _{p=1}^q s_p.
\end{equation*}
We also have from \eqref{eq:squarefrobeniusD} that the $\textbf{z}$ vector entering into the composition of $\hat{Z}$, which defines — together with $\hat{X}$ and $\hat{Y}$ — the elements of the relaxed one-body reduced difference density matrix, has the following property:
$$\textbf{z}\tvec\textbf{z} =\sum _{p=1}^q s_p^2. $$
\begin{theorem}\label{prop:Z2}
Let $m$ and $n$ be two natural integers such that $1\leq m < n$. Let {\normalfont{\textbf{D}}} be an $m\times (n-m)$ matrix. Let {\normalfont{$\textbf{A}_a$}} be an $n\times n$ Hermitian matrix partitioned as
\begin{eqnarray*}{\normalfont{
\textbf{A}_a \coloneqq 
 \left(
  \begin{array}{ll}
   \textbf{0}_m      &  \textbf{D} \\
   \textbf{D}^\dag &  \textbf{0}_{(n-m)}  \\
  \end{array} 
 \right).}}
\end{eqnarray*}
Then,
{\normalfont{\begin{align*}
\forall i \in \llbracket 1, (n-m)\rrbracket, \,\lambda _i ^\downarrow (\textbf{A}_a) &\geq 0, \\
\forall i \in \llbracket 1,m\rrbracket, \,\lambda _{n-m+i} ^\downarrow (\textbf{A}_a) &\leq 0.
\end{align*}}}
\end{theorem}
\begin{proof}
Consider \eqref{eq:eigdecD1} and \eqref{eq:eigdecD2}. Defining $\textbf{S}_+^\downarrow \coloneqq \mathrm{diag}(s_1^\downarrow,\ldots,s_q^\downarrow)$ and $\textbf{S}_-^\downarrow \coloneqq -\mathrm{diag}(s_1^\uparrow,\ldots,s_q^\uparrow)$, a case study shows that, if we consider $\bm{\lambda}^\downarrow(\textbf{A}_a)$, i.e., the matrix of eigenvalues of $\textbf{A}_a$ {sorted in decreasing} order, we get the following results:

$\;$

\noindent In \textit{Case 1a} $(m<(n-m))$ and \textit{Case 1b} $(m>(n-m))$,
$$\bm{\lambda}^\downarrow(\textbf{A}_a) = \textbf{S}_+^\downarrow \oplus \textbf{0}_{n-2q}\oplus \textbf{S}_-^\downarrow.$$
In \textit{Case 2} $(m=(n-m))$, the central block, i.e., $\textbf{0}_{n-2q}$, is absent, and we get
$$\bm{\lambda}^\downarrow(\textbf{A}_a) = \textbf{S}_+^\downarrow \oplus \textbf{S}_-^\downarrow.$$
In the three cases the proposition is true.
\end{proof}
\begin{theorem}[Lidskii-Wielandt] \label{theo:lidskii}\textit{Let} ${\normalfont\textbf{A}}$ \textit{and} ${\normalfont\textbf{B}}$ {be two} $n\times n$ \textit{Hermitian matrices. Let} ${\normalfont\textbf{C}}$ \textit{be their sum. Let} $
\bm{\alpha} ^\downarrow = (\alpha _1^\downarrow, \ldots , \alpha _n^\downarrow)$, $\bm{\beta}^\downarrow = (\beta _1^\downarrow, \ldots , \beta _n^\downarrow)$, and $\bm{\gamma}^\downarrow = (\gamma _1^\downarrow, \ldots , \gamma _n^\downarrow)$ \textit{be the decreasing-order $n$--tuples of the eigenvalues of} {\normalfont$\textbf{A}$, $\textbf{B}$}, \textit{and} {\normalfont\textbf{C}} \textit{respectively. Then, for any choice of} $1 \leq i_1 < \cdots < i_k \leq n$,
$$\sum _{j=1}^k \left(\gamma ^\downarrow _{i_j}- \alpha ^\downarrow _{i_j}\right) \leq \sum _{j=1}^k \beta ^\downarrow _j.  $$
\end{theorem}
\noindent A proof has been reported in section III.4 (theorem III.4.1) of Ref. \cite{bhatia_matrix_1997}. One direct corollary to the Lidskii-Wielandt theorem is
\begin{corollary}\label{cor:lidskii} \textit{Let} ${\normalfont\textbf{A}}$, {\normalfont\textbf{B}}, {\normalfont\textbf{C}}, $\bm{\alpha} ^\downarrow$, $\bm{\beta} ^\downarrow$, and $\bm{\gamma} ^\downarrow$ \textit{be the matrices and $n$-tuples from theorem {\normalfont \ref{theo:lidskii}}. Then, the} $\bm{\gamma}^\downarrow$ $n$--\textit{tuple is majorized by} $\left(\bm{\alpha}^\downarrow + \bm{\beta}^\downarrow \right)$, i.e.,\normalfont
$$
\forall k \in \llbracket 1,n\rrbracket, \, \sum _{j=1}^k \gamma ^\downarrow _j\leq \sum _{j=1}^k \left(\alpha ^\downarrow _j + \beta^\downarrow _j\right)
$$ 
and, in particular, we see that when $k=n$ we have
$$\sum _{j=1}^n \gamma ^\downarrow _j =  \sum _{j=1}^n \left(\alpha ^\downarrow _j + \beta^\downarrow _j\right).$$
\end{corollary}
\begin{proposition}
$(\vartheta + \vartheta ^Z)$ is an upper bound to the value of the charge transferred during a molecular electronic transition.
\end{proposition}
\begin{proof}
\noindent Choosing $\textbf{A} = \bm{\gamma}^\Delta$, $\textbf{B} = \bm{\gamma}^{\mathrm{Z}}$, $\textbf{C} = \bm{\gamma}^\Delta_{rlx}$, and setting $(k=L-N)$ in Corollary \ref{cor:lidskii} gives, together with \eqref{eq:lowerbound} and Theorem \ref{prop:Z2} with ($\textbf{D} = \textbf{Z}$),
$$\vartheta \leq \vartheta ^{rlx} \leq \vartheta + \vartheta ^Z,$$
hence, according to \eqref{eq:chiphiR}, we finally derive an inequality about the quantity of charge transferred during a molecular electronic transition
$$0 \leq q^\mathrm{CT}_{rlx} \leq \vartheta + \vartheta^\mathrm{Z},$$
which is our final result.
\end{proof}


\section{Conclusion}
 We have first revisited, proved and generalized a long-lived conjecture related to the relationship between the relaxed and unrelaxed detachment/attachment density integral values in certain methods. The original conjecture states that in the configuration interaction singles method, the unrelaxed detachment/attachment density integral value is a lower bound to the value of the relaxed one. Its proof was deduced from a first more general result, that we introduced and proved, which holds for the time-dependent Hartree-Fock theory, the time-dependent density-functional theory, and the Bethe-Salpeter equation. An alternative derivation of this first result is also proposed, using inertia of Hermitian matrices. The two derivation paths actually complete each other, and their combination allows to say more about the matrices that are of interest to us in this contribution and beyond this contribution.
 
 Then, we have applied the detachment/attachment picture to the basis relaxation transformation, picturing such a transformation using a departure/arrival situation, where the basis-unrelaxed excited state is the departure state, and the basis-relaxed excited state is the arrival state. The detachment/attachment density matrices related to the unrelaxed-to-relaxed passage were derived, and their trace was shown to be simply equal to the sum of the singular values of the basis-relaxation matrix. 
 
 Finally, we have combined these results for deriving the upper bound to the exact value of the charge transferred during a light-induced molecular electronic transition as the sum of the unrelaxed detachment/attachment density matrix trace and the basis-relaxation detachment/attachment density matrix trace. The final inequality shows that the excitation and relaxation processes may not result in simply additive contributions to the fully relaxed detachment/attachment picture of the electronic state transition. This boundary value can actually be calculated using at most two matrix trace-computing operations and one SVD operation. 
 \newpage
 \section{Appendices}
\subsection{Dual of Theorem \ref{theo:eigembI}}\label{app:dualeigembI}
\noindent \begin{theorem}[Eigenvalue embedding I, dual]\label{theo:eigembdualI} \textit{Let} ${\normalfont\textbf{A}}$ \textit{be an} $n\times n$ \textit{Hermitian matrix partitioned as }
$${\normalfont\textbf{A}} = \left( \begin{array}{cc}
{\normalfont\textbf{B}}^{\textcolor{white}{*}} & {\normalfont\textbf{D}} \\
{\normalfont\textbf{D}}^\dag & {\normalfont\textbf{C}}
\end{array}\right)
$$
\textit{with} ${\normalfont\textbf{B}} \in \mathbb{C}^{m \times m}$, \textit{assuming that} $(m < n)$. \textit{Let} $\bm{\alpha}^\downarrow = (\alpha ^\downarrow _1,\ldots , \alpha^\downarrow _n)$ \textit{and} $\bm{\beta}^\downarrow = (\beta ^\downarrow _1,\ldots , \beta^\downarrow _{m})$ \textit{be the} $n$\textit{--tuple of the eigenvalues of} ${\normalfont\textbf{A}}$ \textit{sorted in the decreasing order and the} $m$\textit{--tuple of the eigenvalues of }${\normalfont\textbf{B}}$ \textit{sorted in the decreasing order, respectively. Then,}
$$\normalfont\forall i \in \llbracket 1, m \rrbracket, \, \alpha _{i}^\downarrow \geq \beta _i ^\downarrow \geq \alpha _{i+n-m}^\downarrow.$$
\end{theorem}  
\begin{proof} 
\noindent Without loss of generality, we introduce $ \textbf{A}_1 \coloneqq -\textbf{A}$, where
$${\normalfont\textbf{A}}_1 \coloneqq \left( \begin{array}{cc}
{\normalfont\textbf{B}}_1^{\textcolor{white}{*}} & {\normalfont\textbf{D}}_1 \\
{\normalfont\textbf{D}}^\dag_1 & {\normalfont\textbf{C}}_1
\end{array}\right).
$$
We define $\bm{\alpha} _1 ^\uparrow$ as the $n$--tuple of eigenvalues of $\textbf{A}_1$ sorted in the increasing order. We also define $\bm{\beta}_1^\uparrow$ as the $m$--tuple of eigenvalues of $\textbf{B}_1$ sorted in the increasing order. We have that 
\begin{align}
-\bm{\alpha}^\downarrow &=  \bm{\alpha}_1^\uparrow,\label{eq:alpha-alpha1A} \\
-\bm{\beta}^\downarrow &=  \bm{\beta}_1^\uparrow.\label{eq:alpha-alpha1B}
\end{align}
For the sake of clarity, we now turn to the following notations: Let $\textbf{M}$ be an $\ell \times \ell$ Hermitian matrix. Then, $(\lambda^\downarrow_i(\textbf{M}))_{i\in\llbracket 1, \ell\rrbracket}$ and $(\lambda^\uparrow_i(\textbf{M}))_{i\in\llbracket 1, \ell\rrbracket}$ are defined as the two $\ell$--tuples of eigenvalues of $\textbf{M}$ sorted in the decreasing and increasing order, respectively. With that in mind we write Theorem \ref{theo:eigembI} for $\textbf{A}_1$, i.e.,
$$\forall i \in \llbracket 1, m \rrbracket, \, \lambda _i^\uparrow (\textbf{A}_1) \leq \lambda _i ^\uparrow (\textbf{B}_1) \leq \lambda _{i+n-m}^\uparrow(\textbf{A}_1).$$
Plugging the definition of $\textbf{A}_1$ and using \eqref{eq:alpha-alpha1A} and \eqref{eq:alpha-alpha1B}, gives
$$\forall i \in \llbracket 1, m \rrbracket, \, -\lambda _i^\downarrow (\textbf{A}) \leq -\lambda _i ^\downarrow (\textbf{B}) \leq -\lambda _{i+n-m}^\downarrow(\textbf{A}),$$
i.e.,
$$\forall i \in \llbracket 1, m \rrbracket, \, \lambda _i^\downarrow (\textbf{A}) \geq \lambda _i ^\downarrow (\textbf{B}) \geq \lambda _{i+n-m}^\downarrow(\textbf{A}),$$
which is the desired result. 
\end{proof}
\subsection{Dual of Theorem \ref{theo:eigembII}}\label{app:dualeigembII}
\noindent \begin{theorem}[Eigenvalue embedding II, dual]\label{theo:eigembdualII} \textit{Let} ${\normalfont\textbf{A}}$ \textit{be an} $n\times n$ \textit{Hermitian matrix partitioned as }
$${\normalfont\textbf{A}} = \left( \begin{array}{cc}
{\normalfont\textbf{B}}^{\textcolor{white}{*}} & {\normalfont\textbf{D}} \\
{\normalfont\textbf{D}}^\dag & {\normalfont\textbf{C}}
\end{array}\right)
$$
\textit{with} ${\normalfont\textbf{C}} \in \mathbb{C}^{(n-m) \times (n-m)}$, \textit{assuming that} $(m < n)$. \textit{Let} $\bm{\alpha}^\uparrow = (\alpha ^\uparrow _1,\ldots , \alpha^\uparrow _n)$ \textit{and} $\bm{\gamma}^\uparrow = (\gamma ^\uparrow _1,\ldots , \gamma^\uparrow _{n-m})$ \textit{be the} $n$\textit{--tuple of the eigenvalues of} ${\normalfont\textbf{A}}$ \textit{sorted in the increasing order and the} $(n-m)$\textit{--tuple of the eigenvalues of }${\normalfont\textbf{C}}$ \textit{sorted in the increasing order, respectively. Then,}
$$\normalfont\forall i \in \llbracket 1,(n-m)\rrbracket, \, \alpha _{i}^\uparrow \leq \gamma _i ^\uparrow \leq \alpha _{i+m}^\uparrow .$$
\end{theorem}  
\begin{proof} 
\noindent Without loss of generality, we introduce $ \textbf{A}_2 \coloneqq -\textbf{A}$, where
$${\normalfont\textbf{A}}_2 \coloneqq \left( \begin{array}{cc}
{\normalfont\textbf{B}}_2^{\textcolor{white}{*}} & {\normalfont\textbf{D}}_2 \\
{\normalfont\textbf{D}}^\dag_2 & {\normalfont\textbf{C}}_2
\end{array}\right).
$$
We define $\bm{\alpha} _2 ^\downarrow$ as the $n$--tuple of eigenvalues of $\textbf{A}_2$ sorted in the decreasing order. We also define $\bm{\gamma}_2^\downarrow$ as the $(n-m)$--tuple of eigenvalues of $\textbf{C}_2$ sorted in the decreasing order. We have that 
\begin{align}
-\bm{\alpha}^\uparrow &=  \bm{\alpha}_2^\downarrow,\label{eq:alpha-alpha2A} \\
-\bm{\gamma}^\uparrow &=  \bm{\gamma}_2^\downarrow.\label{eq:alpha-alpha2B}
\end{align}
Using the notations introduced in the proof of Theorem \ref{theo:eigembdualI} for the tuples of ordered eigenvalues of matrices, we write Theorem \ref{theo:eigembII} for $\textbf{A}_2$, i.e.,
$$\forall i \in \llbracket 1,(n-m)\rrbracket, \, \lambda _i^\downarrow (\textbf{A}_2) \geq \lambda _i ^\downarrow (\textbf{C}_2) \geq \lambda _{i+m}^\downarrow(\textbf{A}_2).$$
Plugging the definition of $\textbf{A}_2$ and using \eqref{eq:alpha-alpha2A} and \eqref{eq:alpha-alpha2B}, gives
$$\forall i \in \llbracket 1,(n-m)\rrbracket, \, -\lambda _i^\uparrow (\textbf{A}) \geq -\lambda _i ^\uparrow (\textbf{C}) \geq -\lambda _{i+m}^\uparrow(\textbf{A}),$$
i.e.,
$$\forall i \in \llbracket 1,(n-m)\rrbracket, \, \lambda _i^\uparrow (\textbf{A}) \leq \lambda _i ^\uparrow (\textbf{C}) \leq \lambda _{i+m}^\uparrow(\textbf{A}),$$
which is the desired result. 
\end{proof}
\noindent An alternative proof is also given in the first step of the alternative proof of Theorem \ref{theo:eigembII} in Appendix \ref{app:eigembII}.

\subsection{Alternative proof of Theorem \ref{theo:eigembII}}
\label{app:eigembII}
\noindent In this appendix we show how to derive Theorem \ref{theo:eigembII} directly from Theorem \ref{theo:eigembI}.
\begin{proof}
Consider the $\textbf{J}$ and $\textbf{A}'$ matrices used in the proof of Theorem \ref{thm:generalizedhaynsworth2}. We see that applying Theorem \ref{theo:eigembI} to $\textbf{A}'$ immediately leads to the dual of Theorem \ref{theo:eigembII}, i.e., Theorem \ref{theo:eigembdualII}. 

Applying Theorem \ref{theo:eigembdualII} to $(-\textbf{A})$ gives, using the notations introduced in the proof of Theorem \ref{theo:eigembdualI} for the tuples of ordered eigenvalues of matrices, 
$$\forall i \in \llbracket 1,(n-m)\rrbracket, \, \lambda_i^\uparrow (-\textbf{A}) \leq \lambda_i ^\uparrow (-\textbf{C}) \leq \lambda_{i+m}^\uparrow (-\textbf{A}).$$
Since we know that 
$$\forall i \in \llbracket 1,(n-m)\rrbracket, \, (\lambda_i^\uparrow (-\textbf{A}) = - \lambda_i^\downarrow (\textbf{A})) \, \land \, (\lambda_i^\uparrow (-\textbf{C}) = - \lambda_i^\downarrow (\textbf{C})),$$
we find that
$$\forall i \in \llbracket 1,(n-m)\rrbracket, \, -\lambda_i^\downarrow (\textbf{A}) \leq - \lambda_i ^\downarrow (\textbf{C}) \leq -\lambda_{i+m}^\downarrow (\textbf{A}),$$
i.e.,
$$\forall i \in \llbracket 1,(n-m)\rrbracket, \, \lambda_i^\downarrow (\textbf{A}) \geq  \lambda_i ^\downarrow (\textbf{C}) \geq\lambda_{i+m}^\downarrow (\textbf{A}),$$
which is the desired result.
\end{proof}

\section{Affiliation of the authors}

Enzo Monino\\
\textit{J. Heyrovsk\'{y} Institute of Physical Chemistry, Academy of Sciences of the Czech \mbox{Republic,}} v.v.i. \textit{Dolej\v{s}kova 3, 18223 Prague 8, Czech Republic}

$\;$

\noindent Jérémy Morere and Thibaud Etienne\footnote{thibaud.etienne@univ-lorraine.fr}\\
{\textit{Université de Lorraine,} CNRS, LPCT, \textit{F-54000 Nancy, France}}\\

\bibliographystyle{unsrt}

\end{document}